\documentclass[
 reprint,
superscriptaddress,
 amsmath,amssymb,
 aps,
]{revtex4-2}

\usepackage{xcolor}
\usepackage{graphicx}
\usepackage{dcolumn}% Align table columns on decimal point
\usepackage{bm}% bold math
\usepackage{float}
\usepackage[abs]{overpic}
\usepackage{lipsum} 
\newcommand{\TM}[1]{\textcolor{red}{\textbf{[TM:}~#1\textbf{]}}}

\begin{document}

\preprint{APS/123-QED}

\title{Stochastic ion emission perturbation mechanisms in atom probe tomography: Linking simulations to experiment\\ }% Force line breaks with \\

\author{Aslam Shaikh}
\email{Corresponding author: aslam.shaikh@aalto.fi}
\affiliation{Department of Applied Physics, Aalto University, P.O. Box 15600, 00076 Aalto, Espoo, Finland}
\author{Tero M\"akinen}
\affiliation{Department of Applied Physics, Aalto University, P.O. Box 15600, 00076 Aalto, Espoo, Finland}
\author{Fran\c cois Vurpillot}
\affiliation{Groupe de Physique des Mat\'eriaux, CNRS, University of Rouen Normandie, 76000 Rouen, France}
\author{Mikko Alava}
\affiliation{Department of Applied Physics, Aalto University, P.O. Box 15600, 00076 Aalto, Espoo, Finland}
\author{Ivan Lomakin}
\affiliation{Department of Applied Physics, Aalto University, P.O. Box 15600, 00076 Aalto, Espoo, Finland}



\begin{abstract}
Field evaporation in atom probe tomography (APT) includes known processes related to surface migration of atoms, such as the so-called roll-up mechanism. They lead to trajectory aberrations and artefacts on the detector.
These processes are usually neglected in simulations.
The inclusion of such processes is crucial for providing reliable models for the development and verification of APT reconstruction algorithms, a key part of the whole methodology.
Here we include stochastic lateral velocity perturbations and a roll-up mechanism to simulations performed using the Robin--Rolland model.
By comparing with experimental data from Al and Ni systems, we find the stochastic perturbation energy distributions that allow us to very accurately reproduce the detector patterns seen experimentally and thus greatly improve the accuracy of the simulations.
We also explore the possible causes of remaining discrepancies between the experimental and simulated detector patterns.

\end{abstract}

\maketitle

\section{Introduction}
Atom probe tomography (APT) is a powerful material characterization technique that enables three-dimensional compositional mapping of materials with near-atomic resolution~\cite{gault2012atom}, offering unique insights into the material structure at the nanoscale. In a typical experiment, a small (radius of curvature 30-200~nm) needle-shaped specimen is subjected to an intense electric field~($\sim 10^{10} \, \text{Vm}^{-1}$) which ionizes surface atoms and causes them to field evaporate. The ions then accelerate towards a time-resolved position-sensitive detector, and their time-of-flight and impact positions are used to reconstruct the material in three dimensions. 

While APT has established itself as a powerful tool, the technique is limited by trajectory aberrations and reconstruction artefacts that arise from the complexity of field evaporation.
Surface atoms do not desorb from a perfectly smooth emitter and protruding atomic sites~\cite{lefebvre2016atom, gault2012atom} experience locally enhanced fields~\cite{vurpillot2000trajectory}. These local gradients distort ion trajectories~\cite{vurpillot2018simulation}, giving rise to image artefacts---enhanced and depleted zone lines signalling intense deviations of trajectories in some crystallographic directions~\cite{waugh1975field}---that challenge the metrological accuracy of the method.

The origin of such distortions has been debated for decades. A natural source for these perturbations would be thermal motion, even at cryogenic base temperatures (typically 20–80~K)~\cite{tsong1990atom}.
In laser-pulsed APT this effect is exacerbated by transient surface heating induced by the laser pulse~\cite{gault2010influence}.
However, as noted already by Waugh~\cite{waugh1975field}, the lateral deflections expected from thermal energies alone are small for ions such as Al$^+$, whereas experimental measurements consistently reveal much larger dispersions~\cite{lefebvre2016atom}. The direct use of thermal effects also does not work in explaining any directional effects. This discrepancy indicates that additional mechanisms must be active.
Electrostatic models~\cite{oberdorfer2013full, oberdorfer2015applications, vurpillot1999shape, vurpillot2015modeling, rolland2015meshless} are also not able to reproduce the extent of the dispersions with thermal energies. Additionally, using kinetic Monte Carlo methods~\cite{gruber2011field} to relax the extremal dynamics approximation commonly employed in simulations---by introducing randomness into the choice of the evaporation event---showed little effect at experimental temperatures with the extent of zones line depletions and enhancements being even larger in size.

Early interpretations emphasized ion-optical effects due to emitter geometry~\cite{krishnaswamy1977multilayer}
but over the years, evidence has accumulated that surface diffusion and local rearrangements prior to evaporation contribute significantly to ion trajectory deviations~\cite{wada1984thermally,gault2012impact, ndiaye2023surface, ohnuma2019surface}.
Using density functional theory~(DFT) under high electrostatic fields for~Al, a strong link was established between the electric field strengths required to trigger field evaporation and to significantly facilitate surface diffusion~\cite{sanchez2004field}.
In addition, the migration due to local electric field gradients can be strongly directional~\cite{tsong2001mechanisms,bassett1963thermal,bassett1975surface,
kellogg1978direct, gomer1990diffusion, ehrlich1980surface,tsong1975dissociation,antczak2007jump}.
In particular, the so-called roll-up process~\cite{waugh1976investigations, ohnuma2022first}, where an atom migrates onto a high-field neighboring site---typically crossing onto an atomic terrace---before desorption, has been demonstrated both experimentally~\cite{suchorski1996noble} and by DFT calculations~\cite{ashton2020ab} as an energetically favorable mechanism for field evaporation.

This rolling-up effect adds lateral kinetic energy to emitted ions~\cite{waugh1976investigations}, through a process we refer to as the "slingshot mechanism" in which momentum generated by roll-up displacement and the bond-breaking process is carried over into a transverse velocity as the ions leave the specimen surface. Roll-up is a phenomenological description of a directional lateral motion observed in first-principles studies~\cite{ashton2020ab, ohnuma2022first} of field evaporation.
Importantly, the extent of the roll-up effect is highly material dependent~\cite{ohnuma2022first}. 
It can be understood as a competition between the force of the evaporating atom’s nearest neighbor bonds and the force of the field tugging on the ion~\cite{ashton2020ab}.
It is pronounced in strongly bonded metals such as Ni, Ir, Pt, or Au. In contrast, the effect is weaker in soft metals with weaker surface bonds such as Cu and Al, where desorption maps resemble those predicted by classical electrostatic models~\cite{vurpillot2015modeling}.
One could quantify this by computing a conceptual ratio $\chi = \Delta E_{\rm field} / E_{\rm diff}$ where $\Delta E_{\rm field}$ represents the field-induced lowering of the lateral migration energy barrier and $E_{\rm diff}$ is the intrinsic surface diffusion barrier. Large $\chi$ would then correspond to the roll-up mechanism and a lower $\chi$ to small isotropic perturbations.\\

Existing simulation frameworks~\cite{rolland2015meshless, oberdorfer2013full, parviainen2014molecular, zanuttini2017simulation, sommer2016field} have successfully reproduced many aspects of field evaporation, but they usually omit these stochastic perturbations~\cite{yao2015effects}. As a result, deviations in detector hit patterns remain insufficiently understood. 
The addition of molecular dynamics steps into field evaporation models is thought to be an elegant way to understand these physical mechanisms~\cite{vurpillot2015modeling}.
Recent studies using these approaches---such as TAPSim-MD~\cite{qi2023origin}, an extension to the TAPSim~\cite{oberdorfer2013full} simulation package---have begun to capture roll-up effects at very small tip radii ($<10$~nm) but remain computationally demanding and do not yet provide a material-specific interpretation of the effect~\cite{qi2023origin}.
This would need a complex and time-consuming integration of the ab-initio approach in the molecular dynamics model.\\

This work addresses the problem of incorporating perturbation mechanisms into field evaporation simulations by explicitly studying the stochastic nature of ion emission and comparing the results directly with experimental desorption patterns. Such an approach is essential because the accuracy of APT reconstructions is fundamentally limited by the quality of detector data, which unavoidably contains artefacts introduced during the evaporation process. By quantifying the impact of random lateral velocities and roll-up distortions on ion trajectories, we establish a physically informed framework for field evaporation modeling. The inclusion of these stochastic perturbations represents a critical step toward improved reconstruction algorithms and a more reliable interpretation of APT data.

\section{Methods}

The simulation framework in this work is based on the Robin--Rolland model (RRM)~\cite{rolland2015meshless}.
To connect simulations with experiment and assess the role of stochastic distortion mechanisms, we first compared our simulation results with experimental data for two benchmark systems, Al and Ni. We then extended the simulation framework to incorporate additional sources of trajectory deviations. In particular, we implemented stochastic lateral velocity perturbations and roll-up motion at the moment of ion emission, enabling a direct evaluation of their influence on detector patterns.

\subsection{Experimental methods}
Ni and Al APT specimens were prepared using electrochemical polishing, starting from pure metal wire specimens and using the solutions and conditions reported in Refs.~\cite{larson2013local, lefebvre2016atom, miller1981atom, miller2014atom, gault2021atom}. At the start of the analysis, all specimens possessed a radius of curvature of less than 50~nm. Needle-shaped specimens were analyzed using the voltage pulsing mode on a local electrode atom probe with a direct flight path (LEAP~5000XS). The detector is 8~cm in diameter and positioned in front of the specimen at the distance of about 10~cm. Analysis temperatures were 25~K for the Al specimen and 80~K for the Ni specimen. The pulse fraction was consistently set to 20~\% of the DC voltage across all analyses. During APT specimen evaporation, the detection rate was maintained constant at about 1 per 100 pulses~(1~\%), with no restrictions on the gradually increasing DC voltage to offset the progressive blunting of the specimen radius. The datasets sub-volumes were chosen in regions with small variations of this voltage, ensuring small, gradual and controlled conditions of field evaporation. Even though multihits comprise 17.0~\% of our dataset, we have verified that no dissociation effects~\cite{saxey2011correlated} are present in the multihits and therefore should not be prominent in the normal single-hits either. This means that the experimentally observed trajectory aberrations result from single-ion perturbations. %\AS{Is this okay?}

\begin{figure*}[tbh!]
    \centering
    \includegraphics[width=0.5\linewidth]{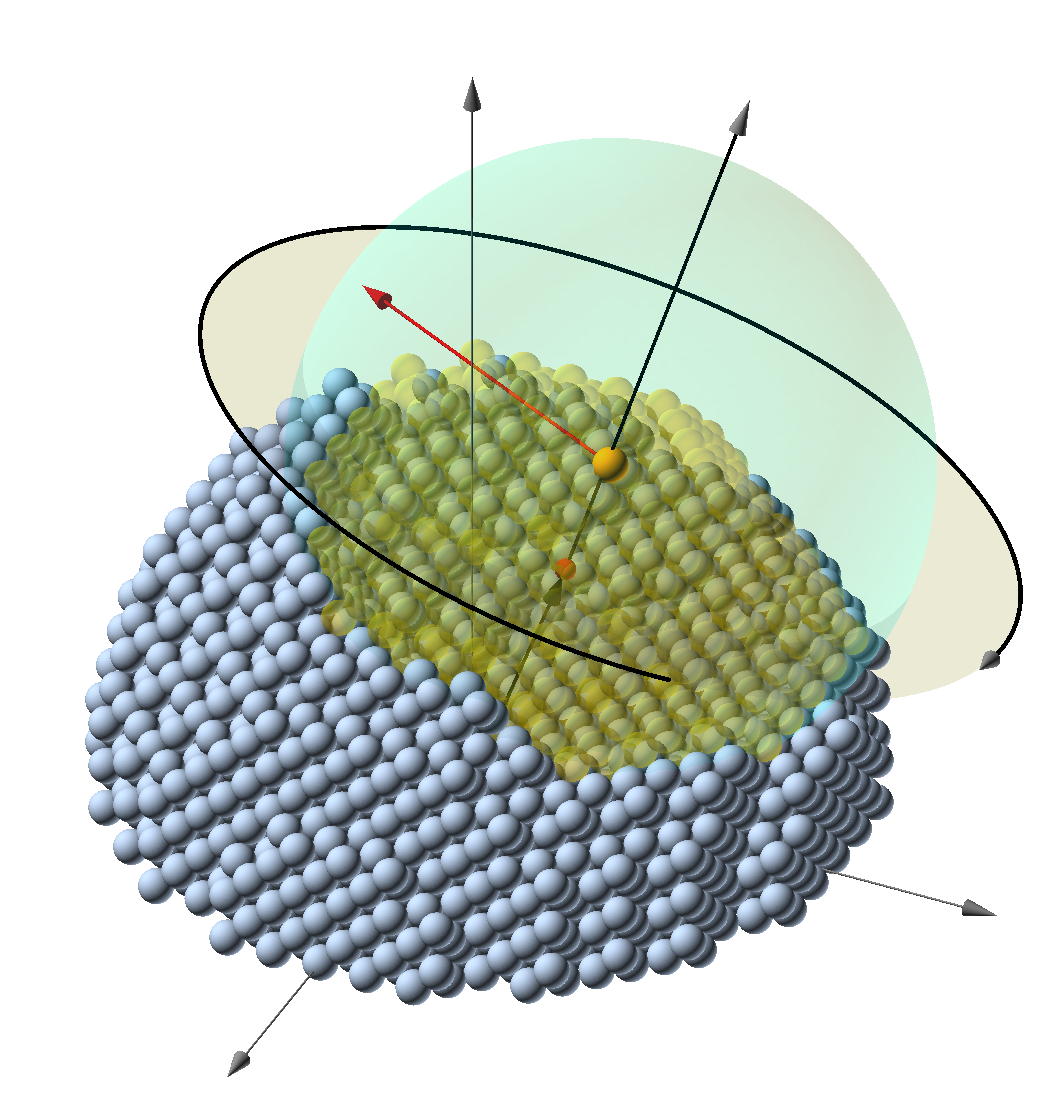}%
        \put(-240,240){\textbf{(a)}}
        \put(-160,235){\Large $\bm{z}$}
        \put(-215,10){\Large $\bm{x}$}
        \put(-30,30){\Large $\bm{y}$}
        \put(-112,112){\Large $\bm{b}$}
        \put(-70,225){\Large $\hat{\bm{n}}$}
        \put(-185,183){\Large \textcolor{red}{$\bm{v}_\perp$}}
    \includegraphics[width=0.5\linewidth]{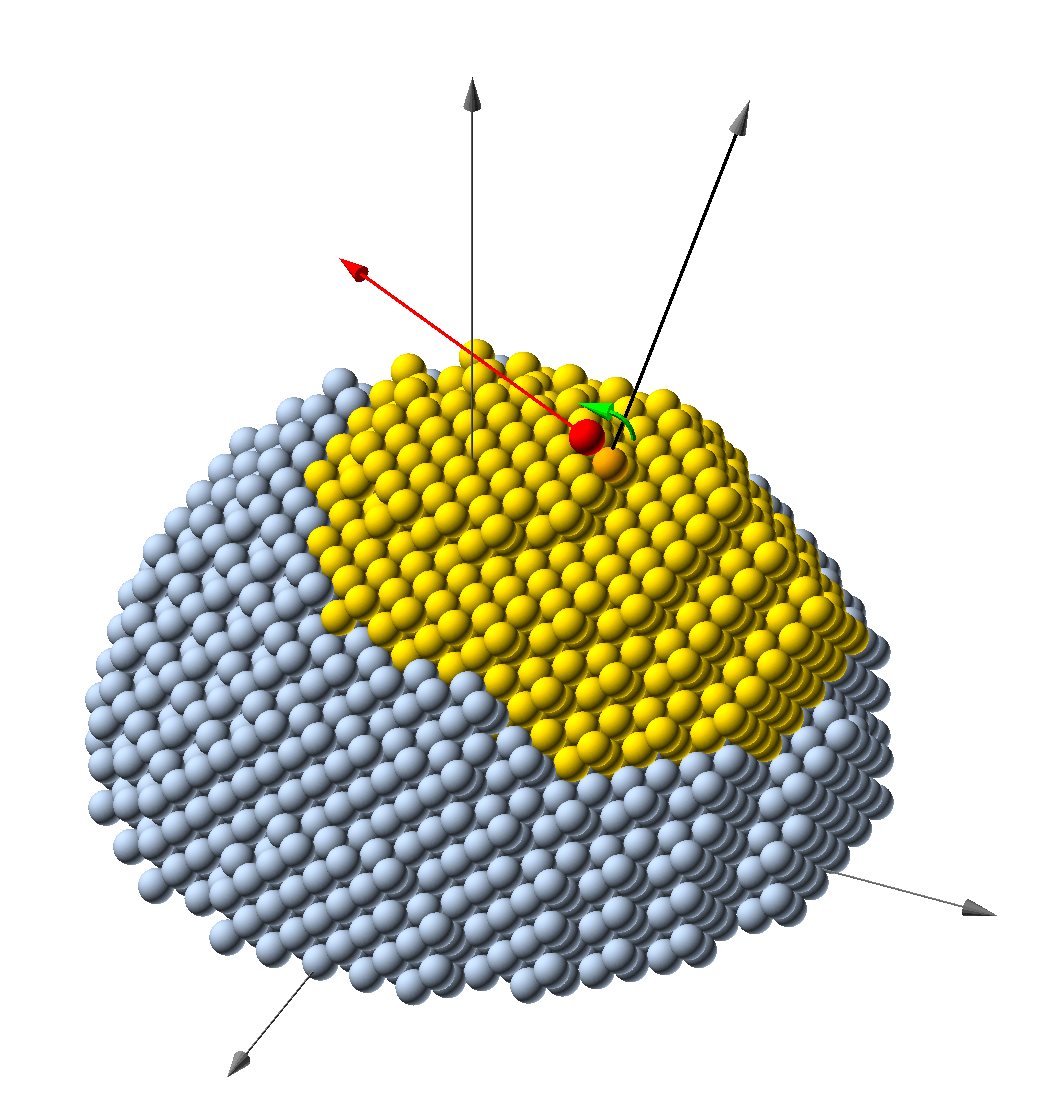}
        \put(-240,240){\textbf{(b)}}
        \put(-160,235){\Large $\bm{z}$}
        \put(-215,10){\Large $\bm{x}$}
        \put(-30,30){\Large $\bm{y}$}
        \put(-70,225){\Large $\hat{\bm{n}}$}  
        \put(-190,185){\Large \textcolor{red}{$\bm{v}_\perp$}} 
    \caption{
    \textbf{(a)}~A schematic drawing of the lateral velocity perturbation for the field evaporation of a fcc hemisphere. The normal vector $\hat{\bm{n}}$ at the evaporation site (orange atom) is obtained by first computing the barycentric vector $\bm{b}$, which runs from the geometric center of the base of the hemispherical cap of the specimen to the center-of-mass (red) of the local neighborhood (green) surrounding the evaporating atom. The surface normal is then defined as the direction from this center-of-mass to the evaporating atom itself. The lateral velocity component $\bm{v}\perp$ is chosen perpendicular to this normal and assigned a random azimuthal direction.
    \textbf{(b)}~A schematic drawing of roll-up distortion implementation where the evaporating atom (orange) is first displaced (rolled-up) on top of the second highest field neighboring atom (red atom and green arrow), and then an additional lateral velocity component $\bm{v}_\perp$ oriented in the roll-up direction and perpendicular to the local surface normal, is then applied.}
    \label{fig:fig1}
\end{figure*}
\subsection{Simulation Methods}
The RRM~\cite{rolland2015meshless} is a meshless approach to field evaporation modeling that leverages Robin's electrostatic formulation~\cite{robin1886distribution}. In contrast to conventional mesh-based methods~\cite{luken2024adapts, xu2015simulation, oberdorfer2013full}, which require discretizing the specimen volume to solve Laplace’s equation, the RRM operates directly on the coordinates of the specimen’s surface atoms. This allows the equilibrium surface charge distribution---and thus the local electric field---to be computed without introducing a volumetric mesh. We summarize here the discrete scheme used in our simulations, and a more detailed description of the underlying electrostatic formulation and its iterative solution are given in Appendix~\ref{app:robin}.\\

In the RRM, %the smooth continuous surface is replaced by the discrete set of atoms on the emitter surface. 
each surface atom is assigned a small surface area $s_{\text{at}}$ and, under an applied potential acquires a surface charge density $\sigma_i$ and a %such that the corresponding 
charge %on atom $i$ is 
$q_i = s_{\text{at}}\,\sigma_i$. Applying Robin's surface-charge formulation to this discrete set of atoms leads to
\begin{equation}
\label{eq:surfaceCharge}
\frac{q_i}{s_{\text{at}}} 
= \frac{1}{2\pi} 
\sum_{\substack{k=1 \\ k \ne i}}^{N} 
q_k 
\frac{\hat{\bm{n}}_i \times \bm{r}_{ik}}{|\bm{r}_{ik}|^3}
\end{equation}
where $\hat{\bm{n}}_i$ is the local outward normal, the distance between the atoms $\bm{r}_{ik} = \bm{r}_i - \bm{r}_k$, and the index $k$ runs over all the $N$ surface atoms. 
The equilibrium charge distribution is obtained by iterating Eq.~\ref{eq:surfaceCharge} for $q_{i,n}$ ($n$ representing the iteration number)
%\begin{equation}
%\label{eq:disc_iter}
%\frac{q_{i,n+1}}{s_{\text{at}}} 
%= \frac{1}{2\pi} 
%\sum_{\substack{k=1 \\ k \ne i}}^{N} 
%q_{k,n} 
%\frac{\hat{\bm{n}}_i \times \bm{r}_{ik}}{|\bm{r}_{ik}|^3}.
%\end{equation}
starting from arbitrary positive initial values $q_{i,0}$.  
For convex surfaces, this sequence converges to the equilibrium charge distribution~\cite{rolland2015meshless}.
To enforce charge conservation in the discrete scheme, the effective atomic surface area $s_{\text{at}}$ is rescaled after each iteration. %such that
%\begin{equation}
%\label{eq:charge_conservation}
%\sum_{i=1}^{N} q_{i,n+1} 
%= \sum_{i=1}^{N} q_{i,n},
%\end{equation}
%as discussed in more detail in Appendix~\ref{app:robin}. 
%The resulting set of charges~$\{q_i\}$ determines the local electric field at each surface atom and serves as input to the subsequent evaporation and ion-trajectory calculations.}

%In RRM, the simulation domain is defined exclusively by the 3D coordinates of the emitter’s constituent atoms. No additional support points are needed. As can be inferred from Eq.~\ref{eq:surfaceCharge}, the iterative sequence can be formulated using only the vectors $\vec{n}_i$ and $\vec{r}_{ik}$, which depend solely on the surface atomic positions. However, it is evident that 
The direct iterative solution of Eq.~\ref{eq:surfaceCharge} entails a computational complexity of $\mathcal{O}(N^2)$, where $N$ represents the number of surface atoms. To mitigate this computational burden, an approach %adopted here draws inspiration from 
based on the Barnes--Hut algorithm~\cite{barnes1986hierarchical} %, originally developed for efficient $n$-body simulations. Barnes–Hut algorithm 
(which hierarchically groups distant atoms and approximates their collective electrostatic contribution) is used. %The algorithm constructs a a tree structure (an octree in three dimensions) that recursively subdivides space into smaller regions until each cell contains only one or a few atoms. 
In this scheme, nearby atoms are treated individually, while distant groups of atoms %(beyond a predetermined threshold distance) 
are aggregated into an equivalent single virtual atom whose position %(given by their center-of-charge) 
and charge %(total charge of all the atoms) 
represents their collective effect. This %approach 
reduces the number of pairwise interactions, %that must be explicity calculated thereby 
lowering the computational complexity %from $\mathcal{O}(N^2)$ 
to $\mathcal{O}(N \log N)$ while maintaining sufficient accuracy% in the calculation of the charge distribution
.

%The identification of surface atoms along with the determination of the local surface normal vectors for these atoms, is a prerequisite for solving~Eq.~\ref{eq:surfaceCharge}. 
Surface atoms are defined as those surrounded by an asymmetric distribution of neighboring atoms. This is identified using a method~\cite{boll2013interpretation}, %To quantify this asymmetry, a method 
%adapted from the work of Boll~\textit{et al.}~\cite{boll2013interpretation},
where the center-of-mass
%was employed. In this approach, a sphere of radius $R$ is centered on each atom within the simulated volume.
%The centroid $\bm{b}$ 
of all atoms located inside the neighborhood of radius~$R$ 
%sphere 
is calculated, and a vector connecting this center-of-mass to the atom is defined. %the vector $\vec{u}_{i}$ %, defined 
%from the centroid to atom's position, is determined. 
The magnitude of this vector clearly separates the %, $d_{i} = |\vec{u}_{i}|$, is compared against a user-defined threshold $\delta$. Atoms for which $d_{i} > \delta$ are classified as 
surface atoms from the bulk ones. %The direction of the vector %$\vec{u}_{i}$ 
%also directly provides the local surface normal $\hat{\bm{n}}$ at that atom’s position.}
%\fixed{The choice of $R$ represents a balance between smoothness in the definition of the surface normal and computational efficiency. A larger $R$ increases the number of atoms within the sphere, thereby improving the accuracy of the local normal, though at the cost of higher computational demand. The user-defined criterion $\delta$ is an ad hoc parameter adjusted to maintain a monolayer of atoms at the specimen’s surface.}
%\fixed{
Once the surface atoms are identified, a parallelized version of Eq.~\ref{eq:surfaceCharge} is applied iteratively. The rapid convergence of the RRM %, initialized from a uniform value of charge, 
enables the calculation of the charge distribution %across the specimen’s virtual surface 
within a few iterations (%convergence criterion $< 10^{-10}$ in fewer 
less than 100). % iterations). %\AS{Does this need more explaining?}. 
The electric field is then computed directly from the charge distribution using Coulomb’s equation (and the Barnes--Hut algorithm). %is used again to achieve computational efficiency.
%The electric field immediately above the surface was obtained directly from the calculated surface charge distribution.

The candidates for field evaporation are evaluated following procedure proposed by Vurpillot \textit{et al.}~\cite{vurpillot2013model} using the charge density or the strength of the electric field (since they are proportional) over all surface atoms. %, and by using the evaporation field of each atom. In this procedure 
Atoms are removed sequentially according to %a local field strength criterion in which the evaporation rate follows 
an Arrhenius-type dependence of the probablity of evaporation on the electric field at the atom. %derived from the Maxwell–Boltzmann distribution, $\Phi_{\mathrm{evap,i}}(t) = \nu_0 \exp(-\beta \Delta E)$, where $\nu_0$ denotes pre-exponential or attempt frequency factor, $\beta = 1/(k_B T)$ is the inverse thermal energy, $F_i$ is the electric field acting on the atom and $\Delta E =Q(F_i)$ represents the energy barrier that depends on the local field. This barrier decreases as the field increases, and a simple expression for it is
%\begin{equation}
%Q(F_i) = C_{A_i}\left(1 - \frac{F_i}{F_{A_i}}\right)
%\end{equation}
%The constant $C_{A_i}$ (about 1~eV for most elements) reflects the binding energy of the atom to the surface, while $F_{A_i}$ is the characteristic evaporation field for that species. The evaporation probability therefore depends strongly on the ratio $F_i/F_{A_i}$ such that atoms experiencing the largest local fields have the highest likelihood of evaporation.}
%\fixed{
To simulate %this process, the evaporation rate $\Phi_{\mathrm{evap}}(t)$ is evaluated for each surface atom during every iteration. The atom to be removed next is selected according to these rates. At very 
the low temperature ($T < 80~\text{K}$) behavior, %the factor $C_{A_i}/k_B T \approx 150$, which means that only atoms with the highest $F_i/F_{A_i}$ values have a significant probability of evaporation. Under these conditions, the field evaporation process becomes almost deterministic---atoms evaporate primarily from the regions of maximum local field on the tip surface.
we use %therefore employ 
the extremal dynamics approximation and always select the atom with the highest evaporation probability (highest electric field) to be evaporated. %}
%\fixed{
%An estimate of $F_{A_i}$ is required in this algorithm. In this simulation, a constant value is assumed for each element, calculated on the basis of fields estimated for pure metals~\cite{lefebvre2016atom}. 
When an atom is selected for evaporation, it is %simply 
removed from the surface, and %the list of surface atoms is updated accordingly. The surface charge of the candidate atom is then redistributed over the remaining atoms so that the total charge of the specimen is conserved. 
the surface charge distribution is %then 
re-evaluated to account for this new configuration (with the total charge conserved). %Because the removal of a single atom changes the charge distribution only slightly, the algorithm converges very rapidly, typically within 5 to 10 iterations.}

The trajectory of the ion toward the detector is determined by solving Newton’s equation of motion, following an approach similar to Ref.~\cite{vurpillot1999trajectories}, by performing the integration using a velocity-Verlet scheme with an adaptive time step. %Ion trajectories are integrated using a velocity-Verlet scheme with an adaptive time step, determined directly by the ion’s instantaneous velocity and acceleration as $\Delta t_{\mathrm{}} = c_{\mathrm{}} \lVert \mathbf{v} \rVert / \lVert \mathbf{a} \rVert$, where $c_{\mathrm{}}$ is a small dimensionless factor. 
In the immediate vicinity of the specimen surface---where the ion has negligible velocity and the electrostatic field varies most strongly---this adaptive procedure produces extremely fine spatial steps %. The first near-surface segments of the trajectory are therefore resolved with sub-nanometer, and in fact sub-picometer, precision 
($\Delta r \approx 10^{-11}~a \text{ to } 10^{-9}~a
$ where $a$ is the lattice parameter). %Such resolution ensures that the strong near-surface field gradients---responsible for the initial ion deflection---are fully resolved. 
As the %ion accelerates away from the emitter and the 
field becomes progressively smoother with distance, the adaptive timestep increases accordingly %and the spatial step length grows correspondingly 
(spatial step size $\Delta r \approx 10^{-8}~a \text{ to } 10^{-4}~a
$). %Farther from the surface, 
Once the ion has moved several nanometers, %away from the apex, the electric field has become sufficiently smooth that 
in the far-field regime, % is reached. In this regime the acceleration varies only weakly, allowing 
much larger timesteps %and correspondingly larger spatial displacements per integration step 
(spatial step size $\Delta r \approx 10^{-4}~a \text{ to } 10^{4}~a
$) can be used without loss of accuracy. %Only a modest number of coarse steps are then required to carry the ion through this quasi-uniform region. 
After the ion has %traversed this far-field zone and 
reached a distance of 8000~$a$ %from the apex---where the field is effectively uniform---
the remainder of the trajectory is linearly extrapolated %analytically 
to the detector~\cite{vurpillot2015modeling}.\\
%This step is considerably simplified in the RRM compared with mesh-based simulation methods which typically require interpolating between mesh points to compute the electric field throughout free space. In contrast, RRM benefits from partial ion modeling, which eliminates the need for such interpolation. 
%The electrostatic force responsible for accelerating the ion is identical to the desorption force acting on atom $i$ given by
% \begin{equation}
% \vec{F}_i = \frac{1}{4\pi\epsilon_0} \sum_{\substack{k=1 \\ k \ne i}}^{N} q_i q_k \frac{\vec{r}_{ik}}{|\vec{r}_{ik}|^3}.
% \end{equation}
%\begin{equation}
%\label{eq:force_discrete}
%\bm{F}_i 
%= \frac{1}{4\pi\varepsilon_0} 
%\sum_{\substack{k=1 \\ k \ne i}}^{N} 
%q_i q_k 
%\frac{\bm{r}_{ik}}{|\bm{r}_{ik}|^3}.
%\end{equation}
%except that the constant term $\bm{r}_{ik}$ now depends on the instantaneous position of the moving ion, $\bm{r}_i$.}\\

Two face-centered cubic (fcc) specimen geometries, oriented along the [012] (matching the experimental Al data) and [111] (for experimental Ni data) crystallographic directions, are simulated %and analyzed 
using the RRM. %field evaporation simulation framework described previously. 
Each specimen consists of a cylindrical shaft (shank angle~$1.5^\circ$) terminated by a hemispherical cap (radius $80a$) extracted from an fcc lattice aligned along the positive $z$-axis. A virtual detector measuring 30~cm $\times$ 30~cm is positioned 10~cm away along the $z$-axis from the specimen apex. 
A static bias voltage of 1000~V is applied to induce the surface charge distribution. At each simulation step, the atom with the maximum local charge---corresponding to the highest local electric field---is selected for evaporation, and the ion trajectory is determined using classical mechanics for a charged particle in an external electrostatic field. %Near the specimen surface, where the electric field gradients are strongest, trajectories are computed with high spatial resolution to precisely capture early-stage trajectory deflections. %Once the ion reaches a distance of 8000~$a$ from the apex (equivalent to 100 tip radii), the field is assumed to be quasi-uniform, and the trajectory is extrapolated analytically to the detector plane to optimize computational efficiency~\cite{vurpillot2015modeling}.
%\textcolor{red}{resolve detector}\TM{if there is some problem here, just write the distances here in the units of $a$, but use the correct value of $a$ (depending on the material) everywhere else}
%\IL{I think this should be described a bit more clear or just put some relevant Ref where this is discussed}
Following each evaporation event, the surface charge distribution is recalculated to reflect the updated geometry, and the next evaporation candidate is selected. Ion positions on the detector are recorded along with other relevant atomic properties for subsequent analysis.

\begin{figure*}[t] 
  \centering
  % Left - (a)
  \begin{minipage}{0.32\textwidth}
    \begin{overpic}[width=\textwidth]{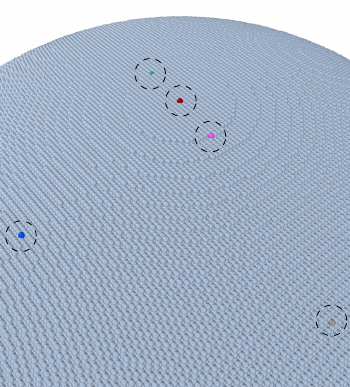}
      \put(5,187){\textbf{(a)}}
    \end{overpic}
  \end{minipage}
  \hfill
  % Middle - (b)
  \begin{minipage}{0.29\textwidth}
    \begin{overpic}[width=\textwidth]{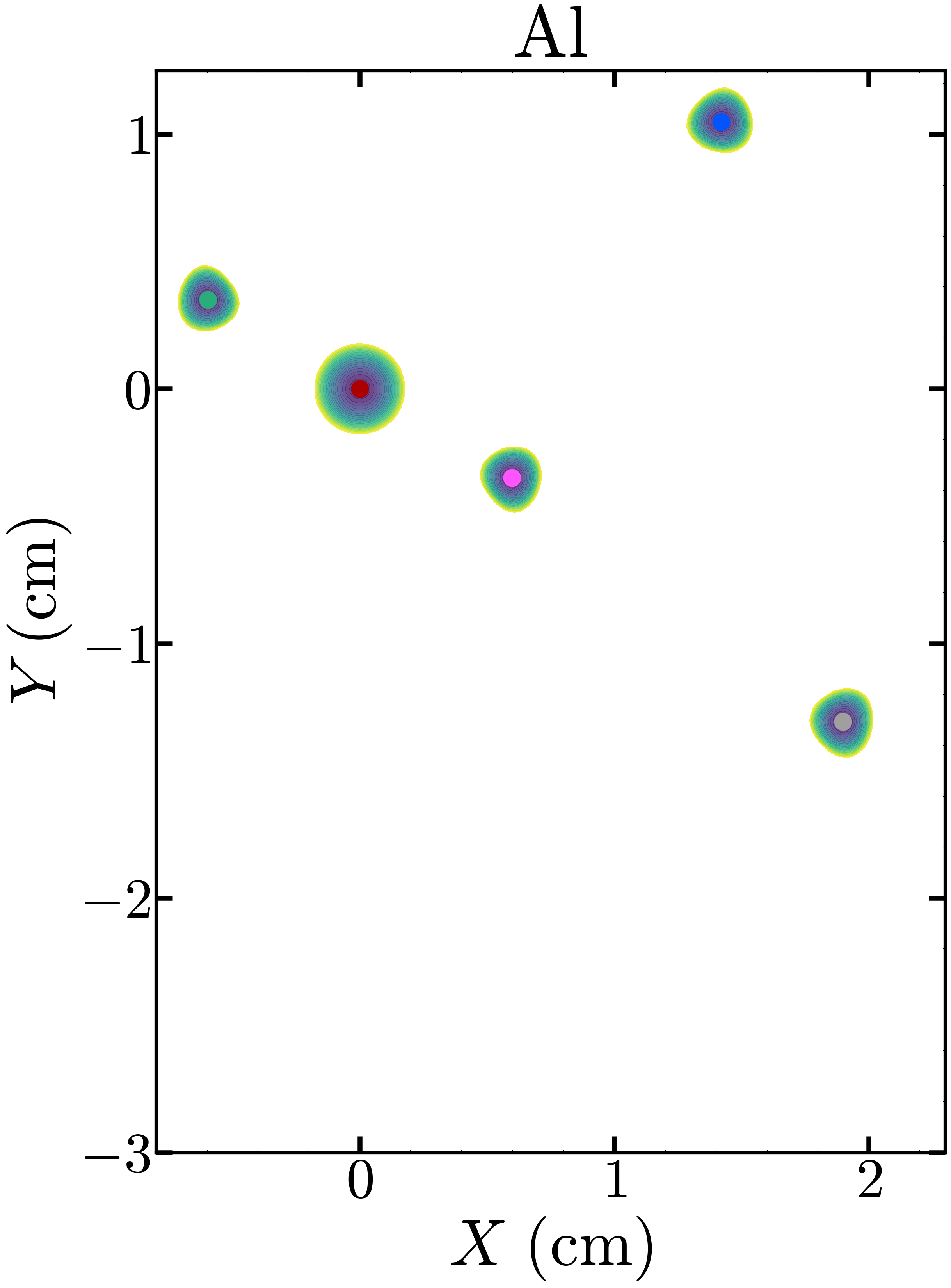}
      \put(5,200){\textbf{(b)}}
      \put(37,28){\includegraphics[width=0.48\textwidth]{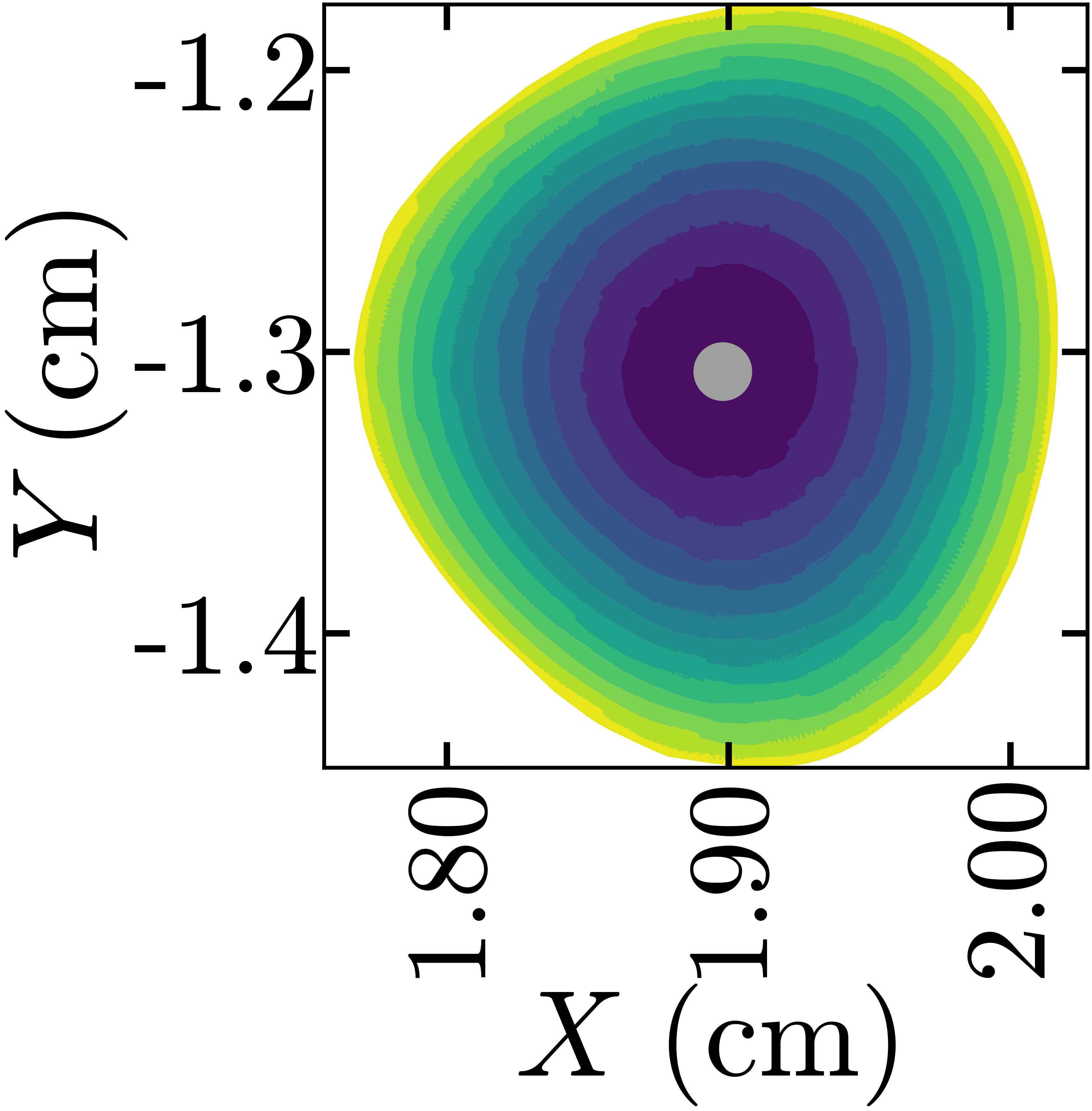}}
    \end{overpic}
  \end{minipage}
  \hfill
  % Right - (c)
  \begin{minipage}{0.37\textwidth}
    \begin{overpic}[width=\textwidth]{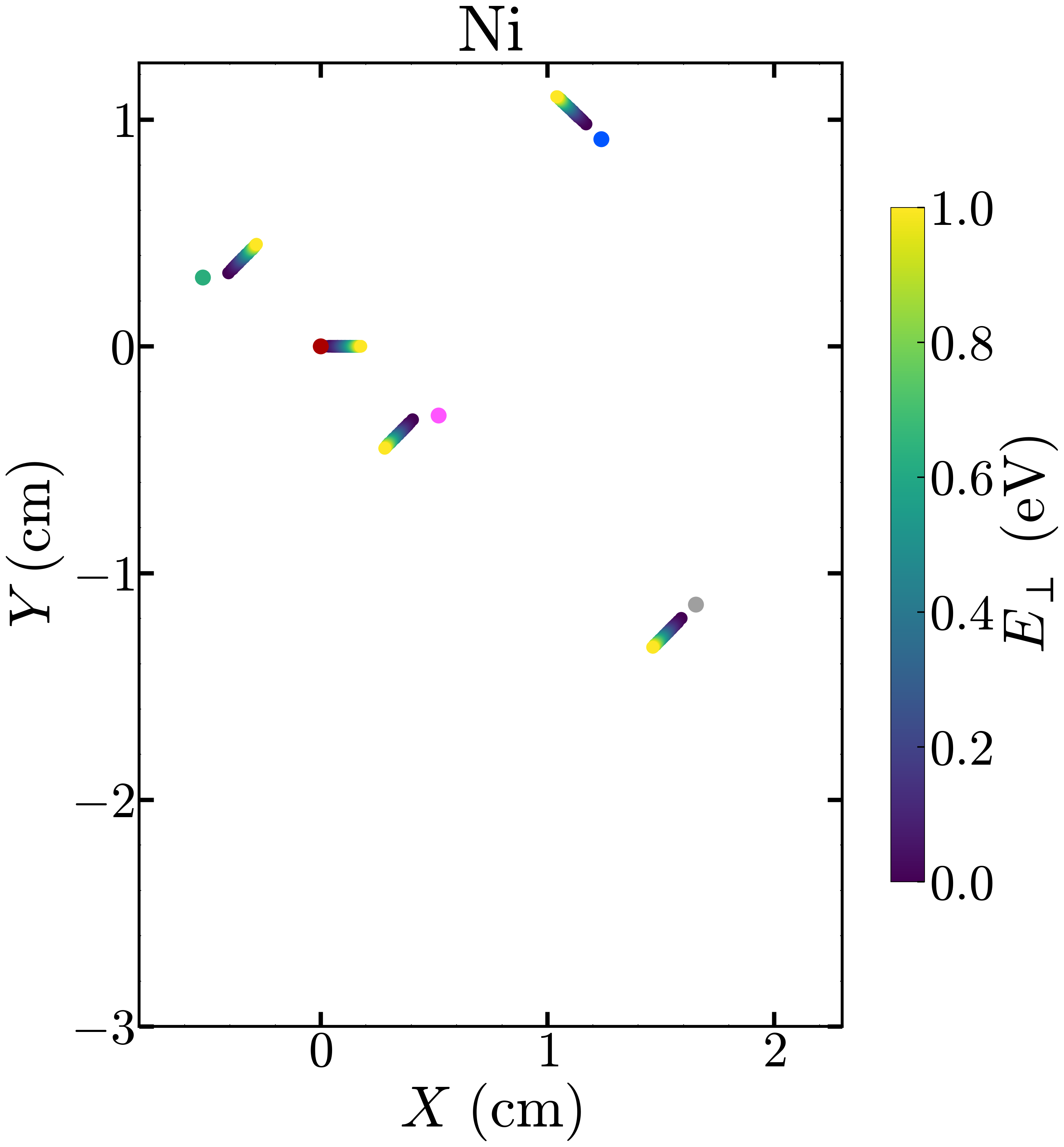}
      \put(5,200){\textbf{(c)}}
      \put(37,28){\includegraphics[width=0.40\textwidth]{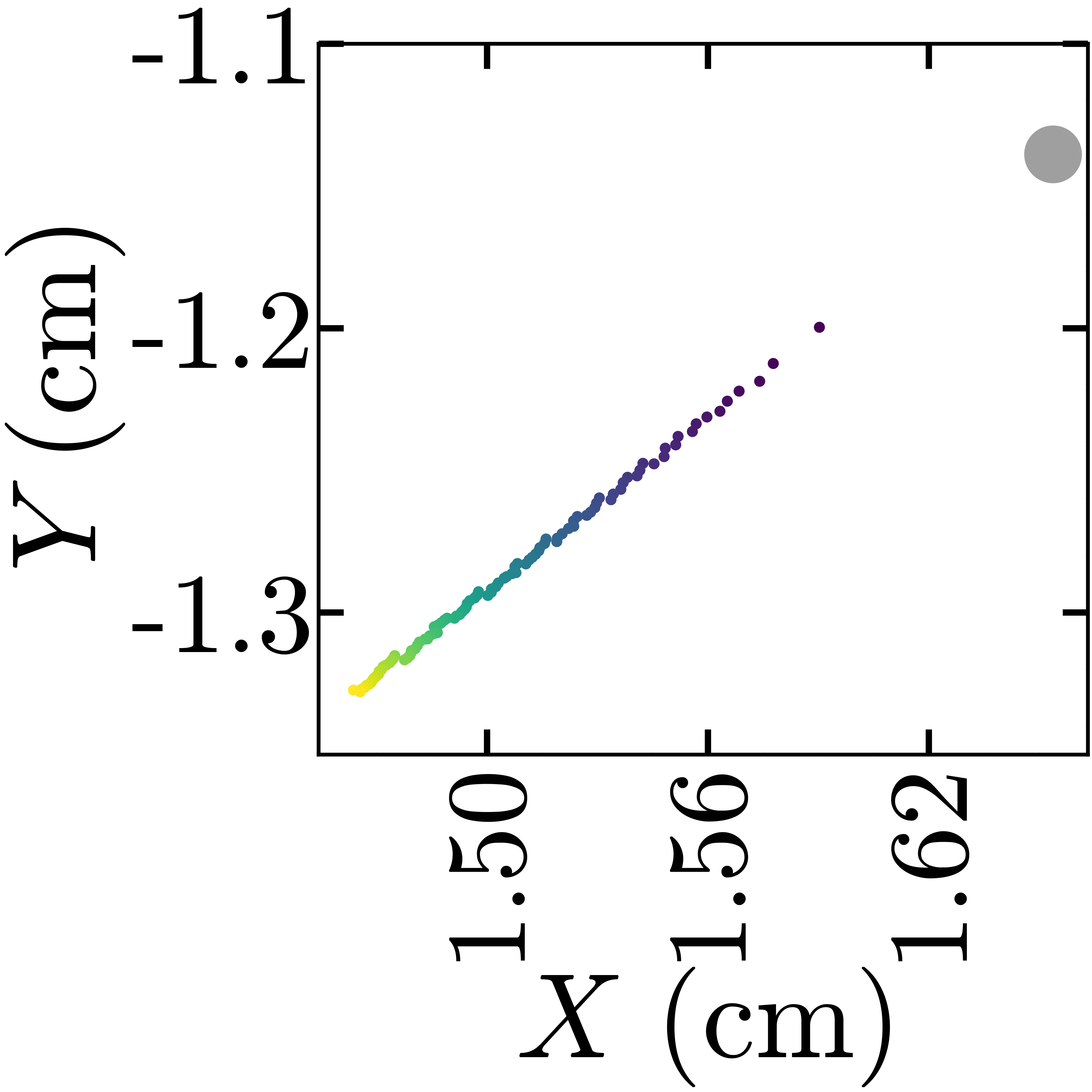}}
    \end{overpic}
  \end{minipage}
  \caption{\textbf{(a)}~Some representative atoms on the sample surface, shown by dashed circles and differentiated by different colors.
  \textbf{(b)}~The effect of lateral velocity perturbation on the detector coordinates $(X, Y)$ for selected evaporated atoms (the ones shown in panel a). The colored points correspond to the detector hit points without any perturbation energy (colors as in panel a) and the colormap shows the perturbation energy corresponding to the other detector hits. All perturbation directions are considered.
  The inset shows a zoomed-in view for one of these atoms.
  \textbf{(c)}~Same as panel b, but for the roll-up perturbation with different energies shown in the colormap. The inset shows the track for a single case.
   }
  \label{fig:NiAl}
\end{figure*}

\subsection{Ion emission perturbations}
Within this RRM framework described above, we have independently incorporated two key distortion mechanisms---initial velocity perturbations and roll-up motion---into the field evaporation simulation.
As Al is an example of a metal with weak surface bonds, the roll-up mechanism is known to be less prominent. Therefore, when simulating the Al system, we have introduced the perturbation mechanism just as an ad-hoc perturbation in the lateral velocity of the ion at evaporation (see~Fig.~\ref{fig:fig1}a). %\AS{It is best to remove this line, commented out} %This perturbation is \fixed{thought} to comprise the "small roll-up" effects. 
The perturbations are introduced by assigning each ion a lateral velocity vector $\bm{v}_\perp$, confined to the plane perpendicular to the local surface unit normal vector $\hat{\mathbf{n}}$ at the ion's position. 
The unit normal is determined by considering a local neighborhood (radius~1.55~$a$) around the evaporating atom, computing its center-of-mass, and constructing the barycentric vector~$\bm{b}$ connecting the geometric center of the hemispherical cap of the specimen to the center-of-mass. The normal vector~$\bm{n}$ is then parallel to $\bm{p} - \bm{b}$ where $\bm{p}$ is the position vector of the ion to be evaporated. This procedure is identical to the normal-definition strategy used in previous field-evaporation simulations based on the RRM~\citep{klaes2021model}, and ensures that the normal reflects both the global curvature of the specimen and local terrace geometry.
The magnitude of $\bm{v}_\perp$ is determined by the specified perturbation energy $E_\perp$ and constitutes the initial velocity of the evaporating ion. The random direction of $\bm{v}_\perp$ is defined by sampling an azimuthal angle from a uniform distribution over the range [$0^\circ$,~$360^\circ$), while the polar angle is fixed at $90^\circ$ with respect to $\mathbf{n}$, ensuring that the velocity lies strictly within the tangential plane. Once the initial velocity is determined, the ion trajectory is then naturally evolved under the influence of the local field $\mathbf{E}$.\\ 

In the case of simulating Ni, a strongly bonded metal, we can implement the full roll-up mechanism where candidate ions first migrate onto the surface site of a neighboring atom exhibiting the highest surface charge prior to evaporation (see Fig.~\ref{fig:fig1}b). This is modeled by repositioning the ion at a new location  $\mathbf{p}_r$, defined as $\mathbf{p}_r = \mathbf{s} + d \hat{\mathbf{n}}$ where $\mathbf{s}$ is the position of the highest-charge nearest neighbor and $d$ is the nearest-neighbor spacing. The roll-up process imparts a lateral velocity $\bm{v_\perp}$, the magnitude of which again depends on perturbation energy $E_\perp$. However, in contrast to Al,  the direction of $\bm{v_\perp}$ is taken to be along the component of the displacement vector $\mathbf{p}_r - \mathbf{p}$ that is perpendicular to the local normal $\mathbf{n}$. 
The roll-up perturbation then combines a small displacement with a lateral velocity component, in contrast with the randomly oriented lateral velocity component of the previous perturbation mechanism.
From the roll-up position $\mathbf{p}_r$, the ion is then projected onto the detector under the effect of the local field $\mathbf{E}$.

\section{Results and Discussion}
To illustrate the effect of the ion emission perturbations, we selected a representative set of atoms occupying different lattice sites---with varying local geometry around different terraces---(see Fig.~\ref{fig:NiAl}a) and applied varying levels of lateral perturbation energy $E_\perp$~($0 - 1~\text{eV}$) when evaporating them. The aberrations seen on detector of course depend on the specific perturbation mechanism. In the case of a stochastic lateral velocity component---corresponding to~Al---increasing $E_\perp$ leads to a nearly radial dispersion of detector positions around the unperturbed reference case ($E_{\perp} = 0$~eV) indicated by the central colored dots in the plots (Fig.~\ref{fig:NiAl}b, colors of the points as in Fig.~\ref{fig:NiAl}a). %\IL{It is not clear that color of ions evaporated ions on Fig.2a and Fig.2b are actually matches}
This radial spread is consistent with the uniform angular distribution of the perturbation applied. The small variations from radially symmetric behavior are related to the variations in the local surface geometry.

In contrast, for the roll-up perturbation---corresponding to~Ni---the direction of the lateral velocity vector is directly determined by the local surface geometry. In the roll-up event, the displacement and the direction of the initial lateral velocity component are given by the position of the highest-charge neighboring atom. 
This can be seen as linear tracks on the detector (Fig.~\ref{fig:NiAl}c) when the perturbation energy is varied. The higher the perturbation energy, the bigger the perturbation is also on the detector. One should note that the jump in the detector positions between the smallest perturbation energies and the unperturbed case is due to the roll-up displacement, which is not included in the unperturbed case.
The magnitude of the change in the detector position due to the displacement (the difference between the unperturbed and smallest $E_\perp$ case) corresponds to a similar change as an initial velocity corresponding to roughly $E_\perp = 0.5$~eV.\\

\subsection{Stochastic perturbation energies}

\begin{figure}[tb!]
    \centering
    \includegraphics[width=\columnwidth]{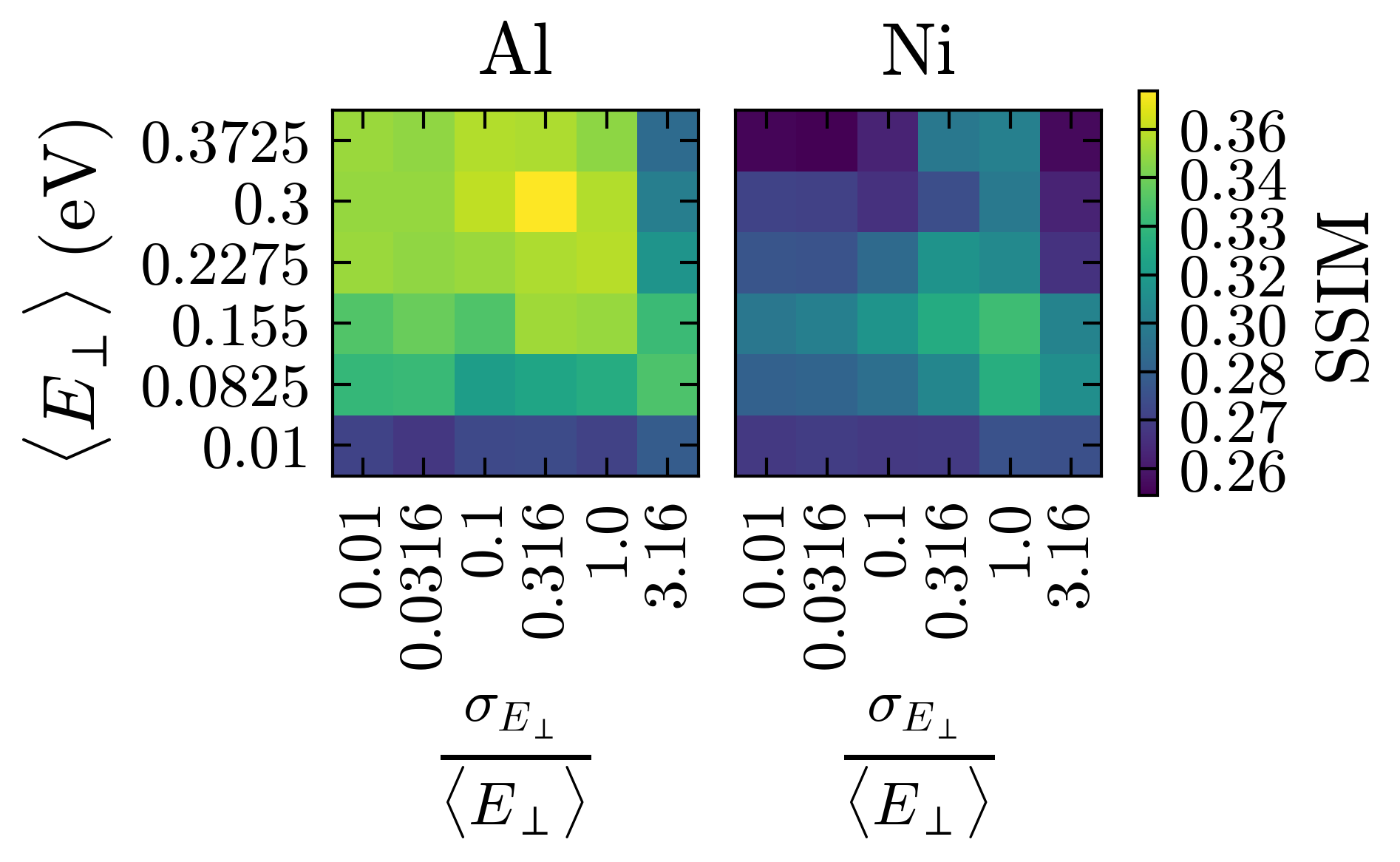} 
    \caption{
    Matrices of the structural similarity index measures~(SSIMs) between the experimental detector images and simulated detector images for the lateral velocity perturbation corresponding to Al experiments (\textbf{left}) and the roll-up perturbation corresponding to Ni experiments (\textbf{right}). The axes correspond to the values of the mean perturbation energy $\langle E_\perp \rangle$ and the standard deviation $\sigma_{E_\perp}$ normalized by the mean. A clear maximum of the SSIM is seen in both cases, corresponding to an optimal distribution of stochastic perturbation energies.}
    \label{fig:ni_ssim_heatmap}
\end{figure}

\begin{figure*}[tb!]
    \centering
    \includegraphics[width=\textwidth]{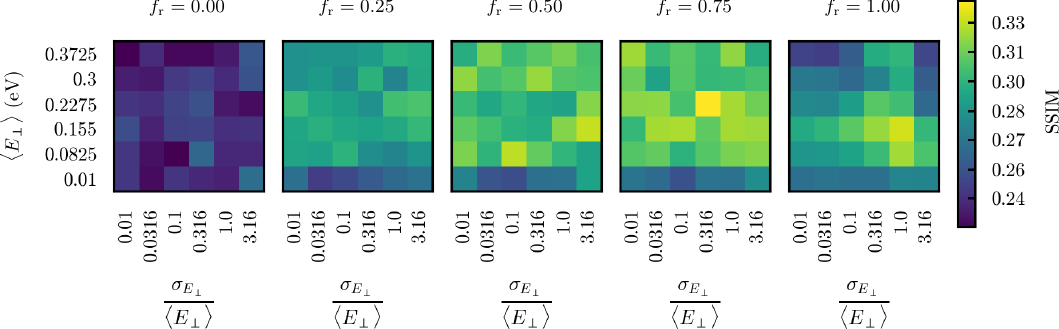} 
    \caption{Matrices of the SSIMs between the experimental detector images and simulated detector images for Ni, when both lateral velocity perturbation and roll-up are considered. The different matrices correspond to different fractions $f_{\rm r}$ of the total perturbation energy $E_\perp$ (mean $\langle E_\perp \rangle$, standard deviation $\sigma_{E_\perp}$) used for the roll-up perturbation.}
    \label{fig:combined_ssim_heatmap}
\end{figure*}

The previous analysis has shown the effect of specific perturbation energies on representative atoms on the sample surface. However, in experimental situations these energies are most likely stochastic quantities depending on a multitude of factors.
To quantitatively assess the influence of these stochastic perturbations, simulations were performed in which perturbation energies~$E_\perp$ for each atom were drawn from truncated normal distributions (only positive values) with an underlying normal distribution with mean $\langle E_\perp \rangle$ and standard deviation $\sigma_{E_\perp}$. The values of the mean ranged from 0.01 to 0.3725 eV, sampled at uniform intervals. The relative standard deviation (RSD), defined as $\sigma_{E_\perp} / \langle E_\perp \rangle$, was varied between 0.01 and 3.16, with logarithmically spaced intervals, to study the effects of very low to very high variances in the energies.
The simulations yield detector maps that can be directly compared with the experimental counterparts. Before the comparison, the simulated maps were cropped to show exactly matching features with the experimental maps.
For this comparison we have used the structural similarity index measure~(SSIM)~\cite{wang2004image} as a quantitative metric. We emphasize that SSIM is used here as a structure-oriented measure to assess whether the simulations reproduce the same characteristic detector-scale features as observed experimentally, rather than to enforce pixel-by-pixel agreement between detector maps. Metrics based on local intensity differences or probability distributions tend to strongly penalize small spatial misalignments and localized discrepancies, which in the present context can arise from experimental–simulation alignment uncertainties, the unknown initial tip shape, and its subsequent evolution during evaporation. At the current level of agreement, such effects are not directly related to the stochastic emission physics studied here and would therefore dominate pixel-level comparison metrics. As the agreement between simulations and experiments improves further and uncertainties in alignment and tip geometry are reduced, more local or distribution-based metrics may become appropriate and complementary.

Based on the SSIMs, we can determine the distributions (i.e.~the mean and standard deviation) that most closely match the experimental data (see Fig.~\ref{fig:ni_ssim_heatmap}).
For Al and simulations with just lateral velocity perturbations, the optimal agreement between experimental and simulated maps was obtained for a normally distributed perturbation energy centered at 0.3~eV and with a standard deviation parameter of 0.0948~eV (RSD~$\sigma_{E_\perp} / \langle E_\perp \rangle = 0.316$), truncated at zero. 
This means that the perturbation energies are very high, with moderate dispersion. With the moderate dispersion the effect of truncation is minimal, giving a true mean of 0.3~eV and standard deviation of 0.0944~eV.
In the case of Ni and simulations with roll-up, the best match is  with normally distributed energies centered at 0.155~eV and with a standard deviation parameter of 0.155~eV (RSD of 1.0), truncated at zero. 
The actual SSIM values are slightly lower than in the Al case.
Here the energies are lower, but the relative dispersion is much higher. This means that the true mean (with the truncation taken into account) is 0.200~eV and the standard deviation 0.123~eV.\\

These values can be compared with theoretical estimates---which constitute lower bounds---for field-induced polarization for small displacements~\cite{baibuz2022polarization} corresponding to scales relevant for roll-up (see Appendix~\ref{app:polar} for full calculation).
These computed values are around half a (true) standard deviation below the true mean of the distributions obtained with the maximization of SSIMs, for both Al and Ni. 
In addition to the estimates being lower bounds, there are a few possible other sources of discrepancy. 
The displacement of the ions might exceed the one lattice step, the local curvature can have large local enhancements (especially near the atomic terraces where roll-up effects are most prominent), and finally, the effective polarizability of near-surface adatoms can be larger than the isolated atom value used in the computation.
We also note that experimental estimates of the energies related to the ion aberration mechanisms~\cite{suchorski1996noble} have indicated energy scales around 0.5~eV---much higher than the thermal energies which are in the range of a few meV---which is not too far from our results, if one considers the fairly broad distributions.\\

\begin{figure*}[tbh!]
  \centering

  % Top row 
  \begin{minipage}{0.32\textwidth}
    \centering
    \begin{overpic}[width=\linewidth]{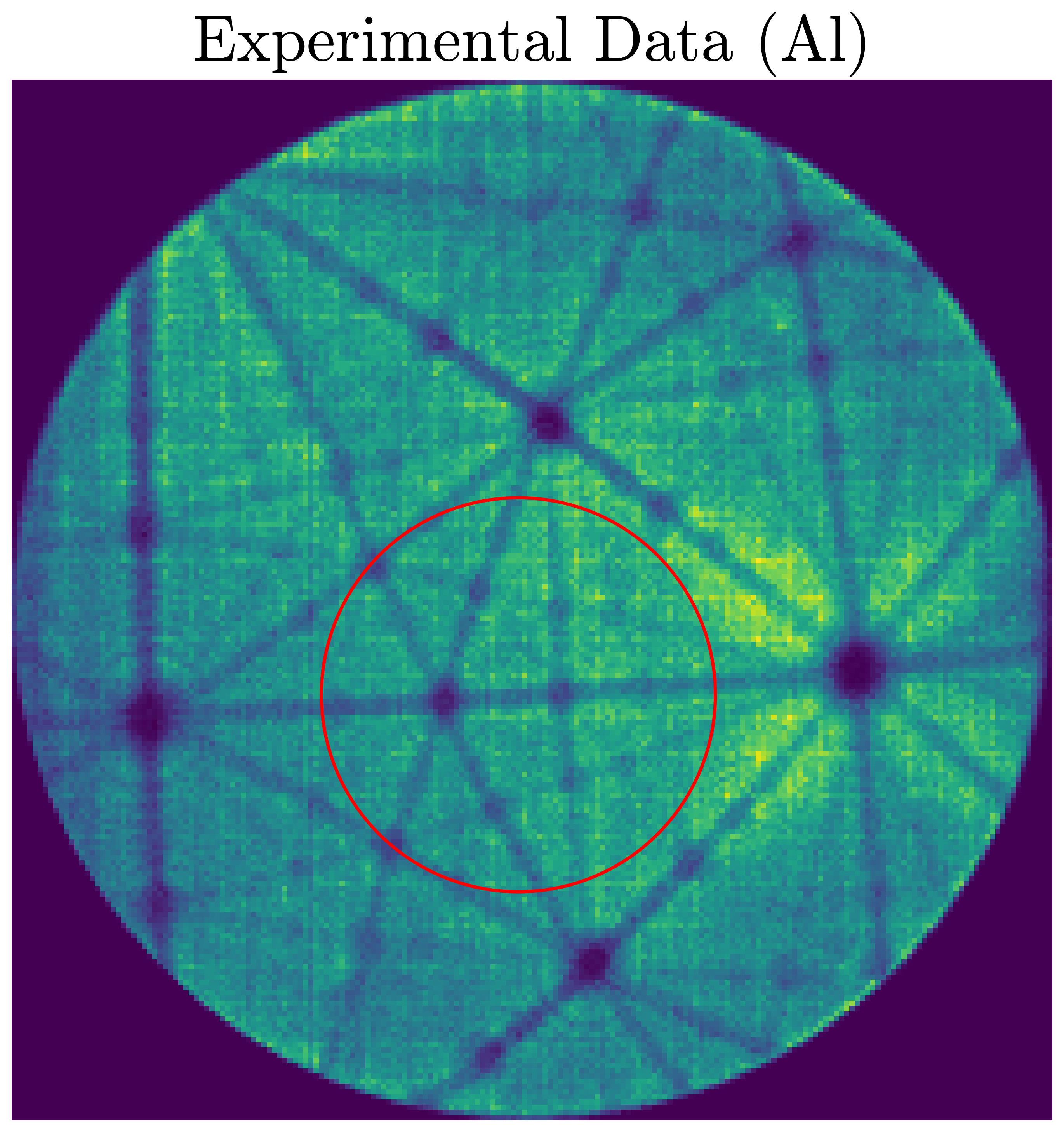}
      \put(3,150){\color{white}\textbf{(a)}}
    \end{overpic}
  \end{minipage}
  \hfill
  \begin{minipage}{0.32\textwidth}
    \centering
    \begin{overpic}[width=\linewidth]{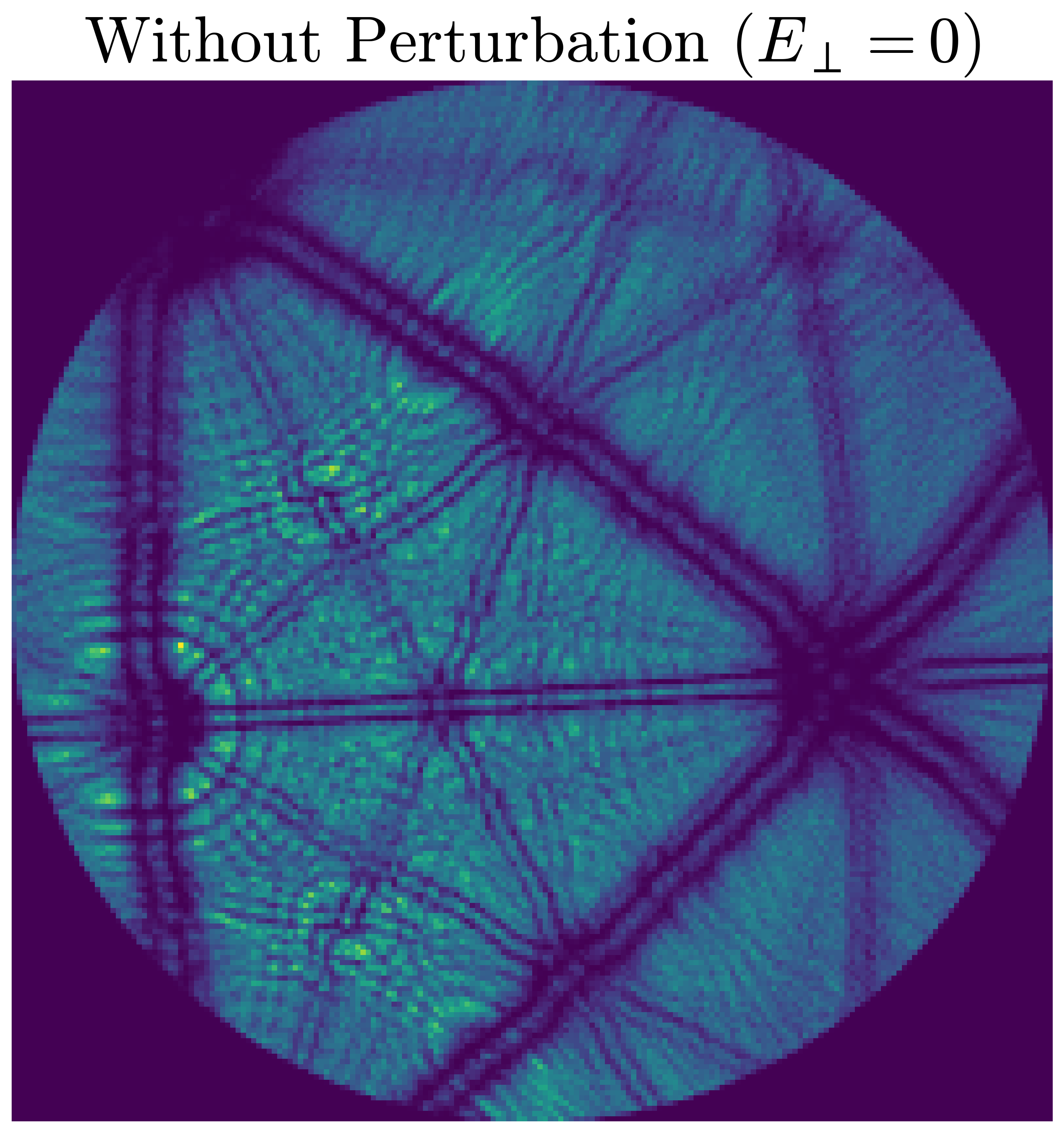}
      \put(3,150){\color{white}\textbf{(b)}}
    \end{overpic}
  \end{minipage}
  \hfill
  \begin{minipage}{0.32\textwidth}
    \centering
    \begin{overpic}[width=\linewidth]{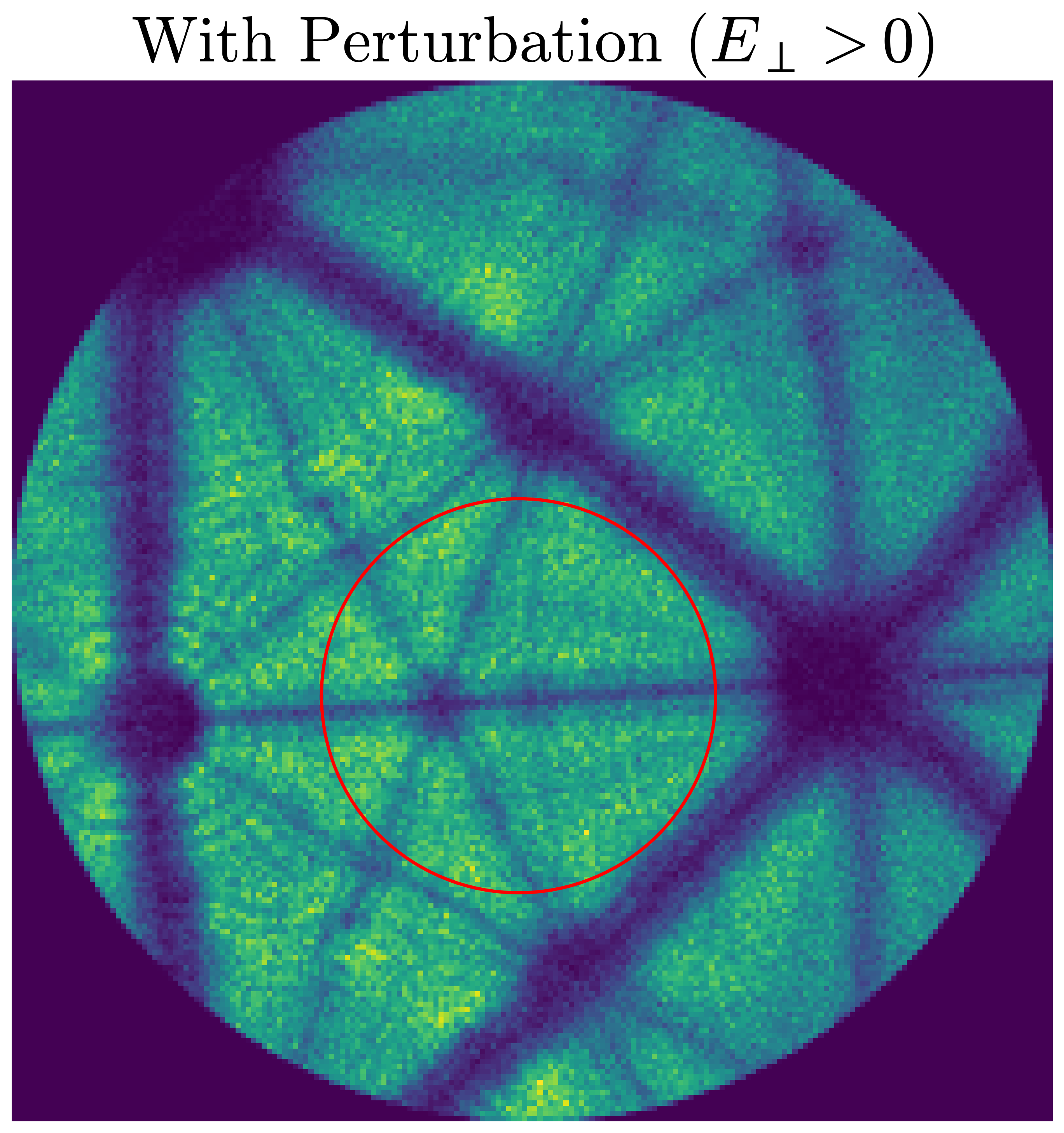}
      \put(3,150){\color{white}\textbf{(c)}}
    \end{overpic}
  \end{minipage}

  \vspace{1em} 

  % Bottom row 
  \begin{minipage}{0.32\textwidth}
    \centering
    \begin{overpic}[width=\linewidth]{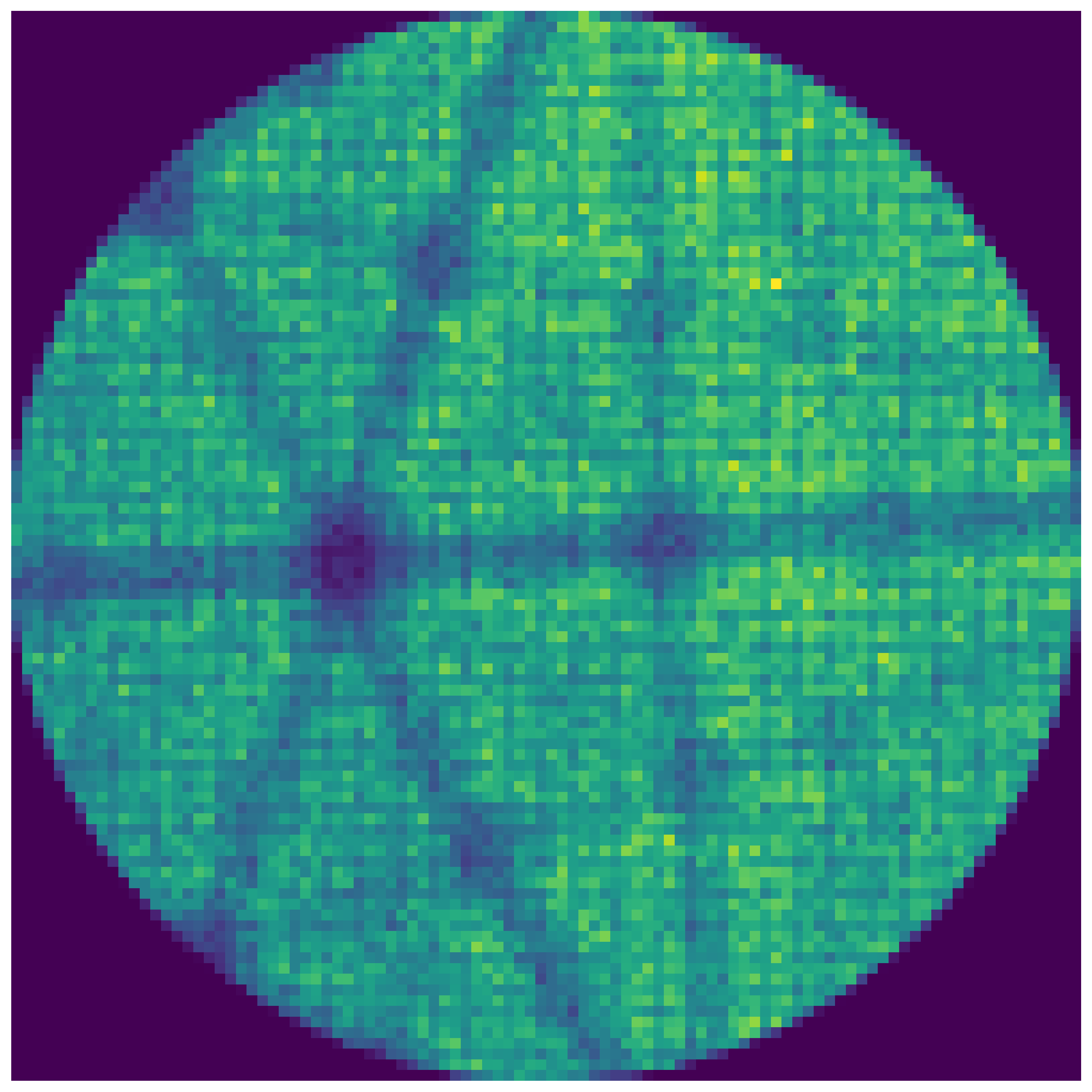}
      \put(3,150){\color{white}\textbf{(d)}}
    \end{overpic}
  \end{minipage}
  \hspace{0.01\textwidth}
  \begin{minipage}{0.32\textwidth}
    \centering
    \begin{overpic}[width=\linewidth]{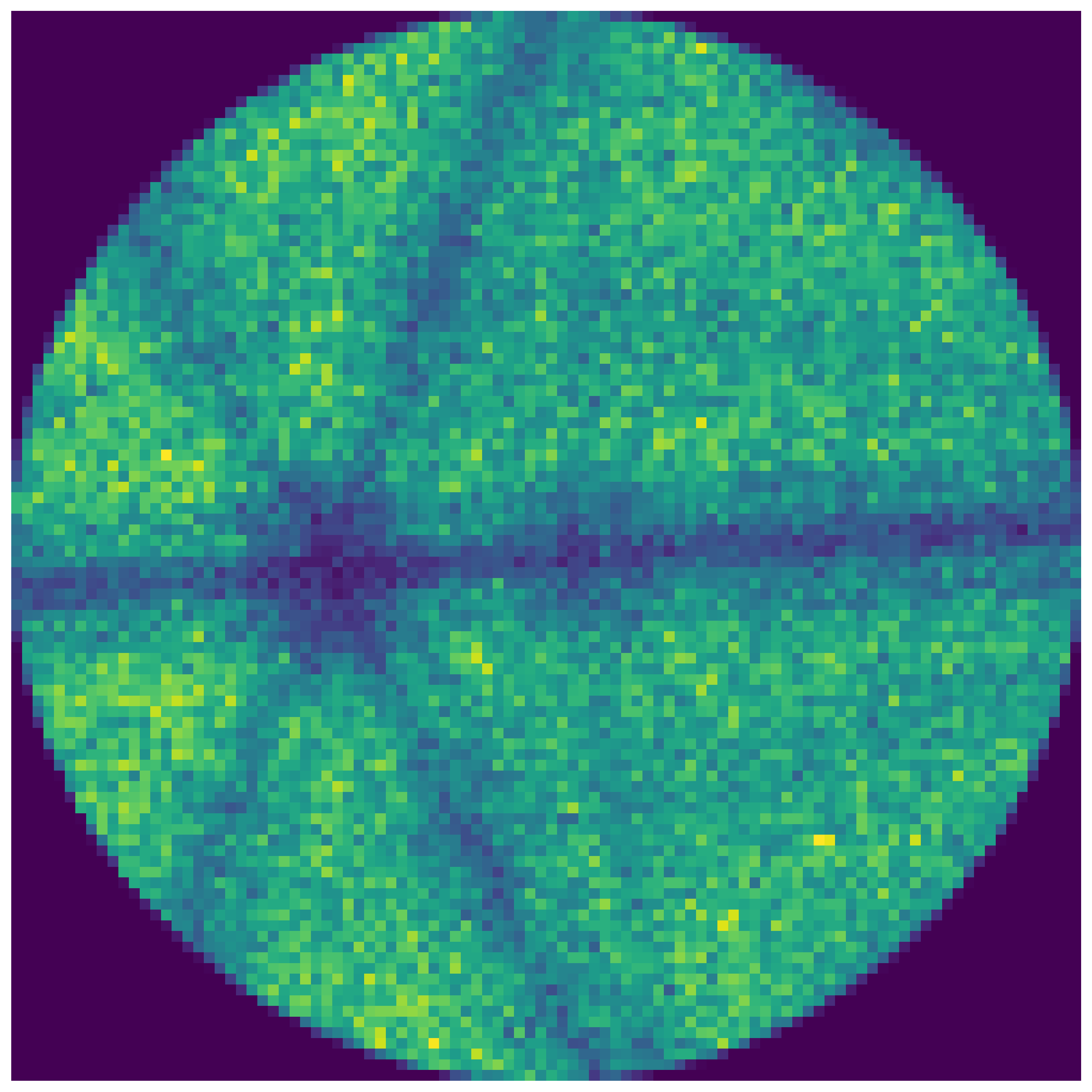}
      \put(3,150){\color{white}\textbf{(e)}}
    \end{overpic}
  \end{minipage}
    \caption{Comparison of experimental and simulated detector maps for the Al[012] system, illustrating the effect of lateral velocity perturbations on detector pattern formation. 
    \textbf{(a)}~The detector maps for experimental Al[012] showing characteristic artefacts including enhanced and depleted zone lines, as well as depleted poles. 
    \textbf{(b)}~The simulated detector map without the added lateral velocity perturbation ($E_\perp = 0$), showing clearly different characteristic features.
    \textbf{(c)}~Simulated detector map with optimal perturbation energy distribution ($E_\perp > 0$), producing patterns that qualitatively resemble the experimental map in panel a.
    To illustrate the similarity, we concentrate on a specific area (red circle in panels a and c) and show the magnified views for \textbf{(d)}~experimental data and \textbf{(e)}~simulation data with perturbations. 
  }
  \label{fig:al_figure_layout}
\end{figure*}

The two mechanisms are not independent, and a natural question is, what would be the effect of combining them? We studied this for the case of Ni, by drawing total perturbation energies~$E_\perp$ from the same truncated normal distributions, but adding another variable~$f_{\rm r}$ which is the fraction of the energy used for roll-up. This then means that the perturbation energy for rollup is $f_{\rm r} E_\perp$ and the perturbation energy for lateral velocity perturbations in a random direction is $(1-f_{\rm r}) E_\perp$. The results are shown in Fig.~\ref{fig:combined_ssim_heatmap}, which is just Fig.~\ref{fig:ni_ssim_heatmap} for different values of $f_{\rm r}$. We see that the optimal similarity with the experimental results is obtained by assigning 25~\% of the perturbation energy to lateral velocity perturbations, but at the same time increasing the mean energy and decreasing the variance of the energies. Therefore the mean roll-up energy is kept roughly the same, but the stochasticity is achieved by perturbations in the velocity direction, instead of variations in the directional lateral velocity.

However, the difference in the SSIM compared to the simple roll-up case ($f_{\rm r} = 1$) is small (also when looking at the detector maps instead of just the SSIM values), and the SSIM value is still lower than in the case of Al. Therefore, for simplicity, we have only considered the $f_{\rm r} = 1$ case in the analysis that follows.

\begin{figure*}[tbh!]
  \centering

  % Top row 
  \begin{minipage}{0.32\textwidth}
    \centering
    \begin{overpic}[width=\linewidth]{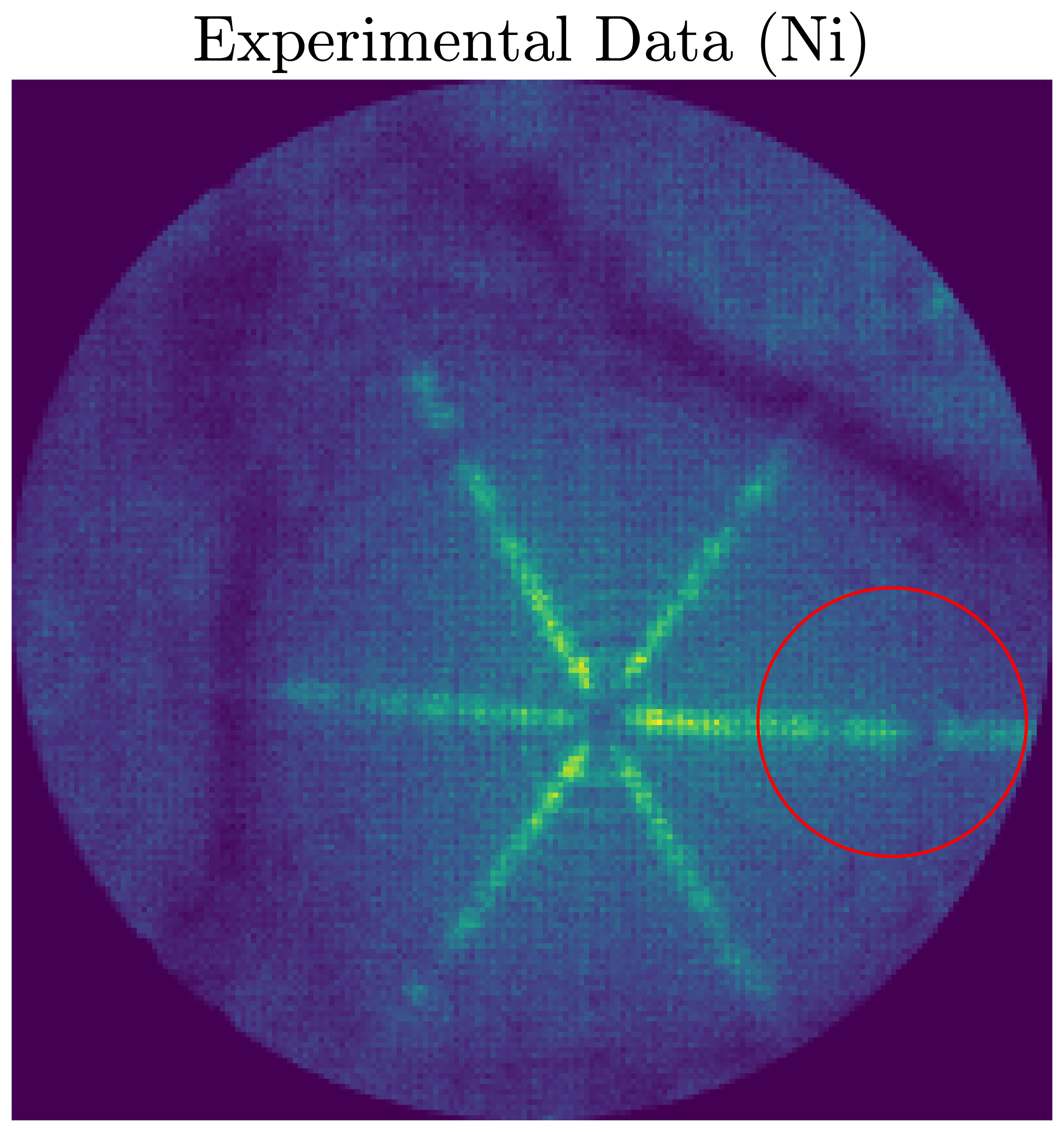}
      \put(3,150){\color{white}\textbf{(a)}}
    \end{overpic}
  \end{minipage}
  \hfill
  \begin{minipage}{0.32\textwidth}
    \centering
    \begin{overpic}[width=\linewidth]{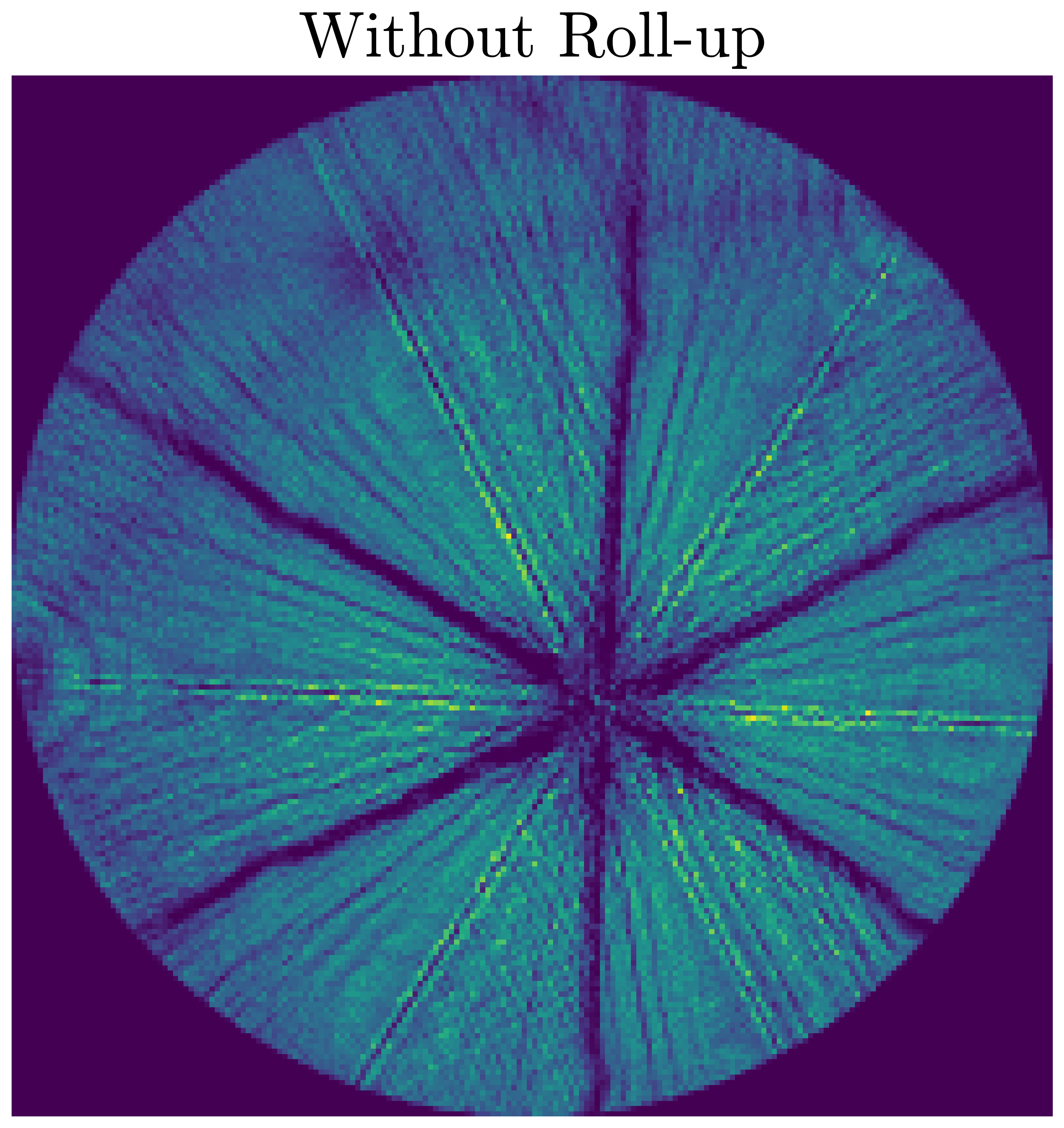}
      \put(3,150){\color{white}\textbf{(b)}}
    \end{overpic}
  \end{minipage}
  \hfill
  \begin{minipage}{0.32\textwidth}
    \centering
    \begin{overpic}[width=\linewidth]{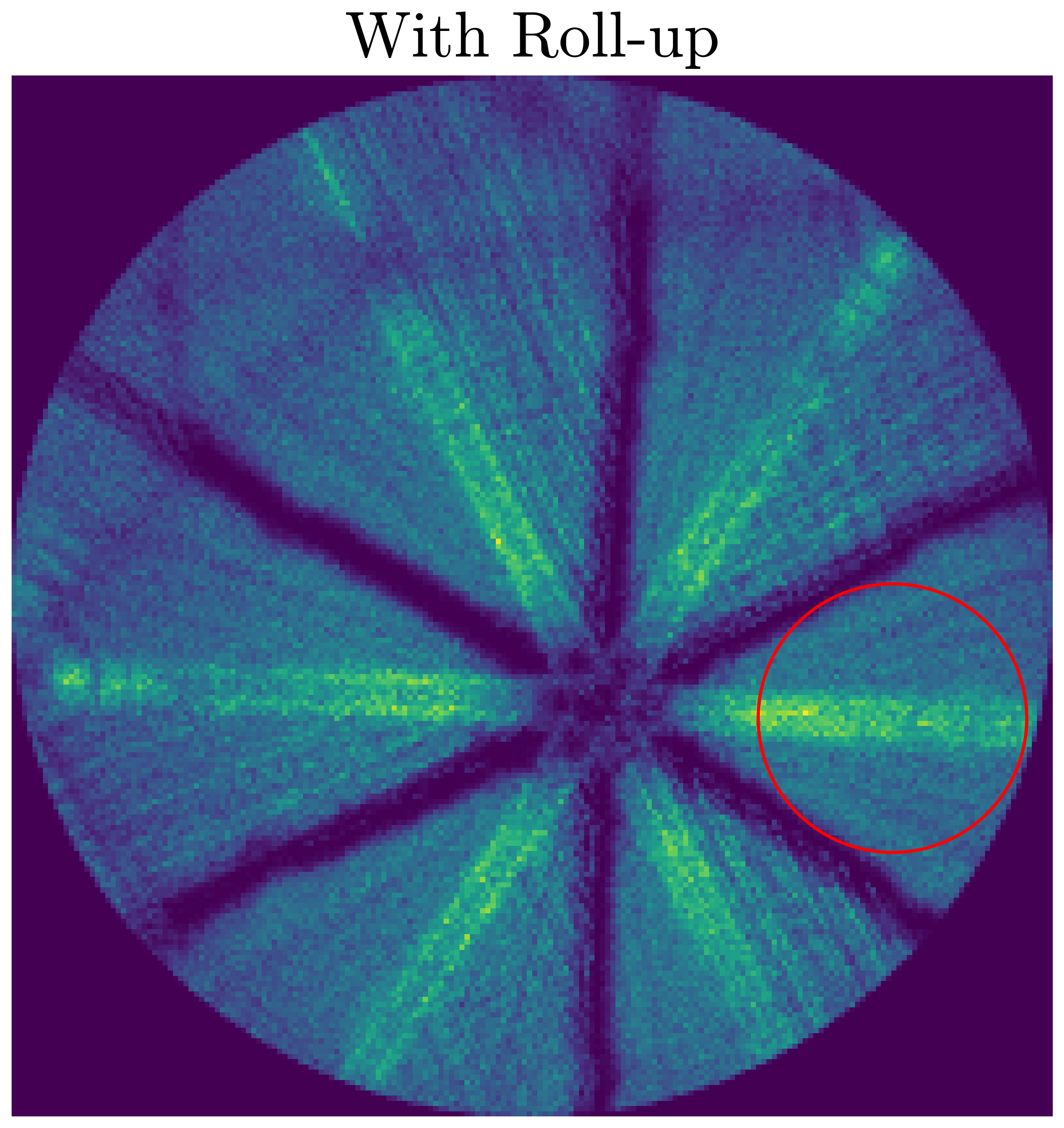}
      \put(3,150){\color{white}\textbf{(c)}}
    \end{overpic}
  \end{minipage}

  \vspace{1em} 

  % Bottom row 
  \begin{minipage}{0.32\textwidth}
    \centering
    \begin{overpic}[width=\linewidth]{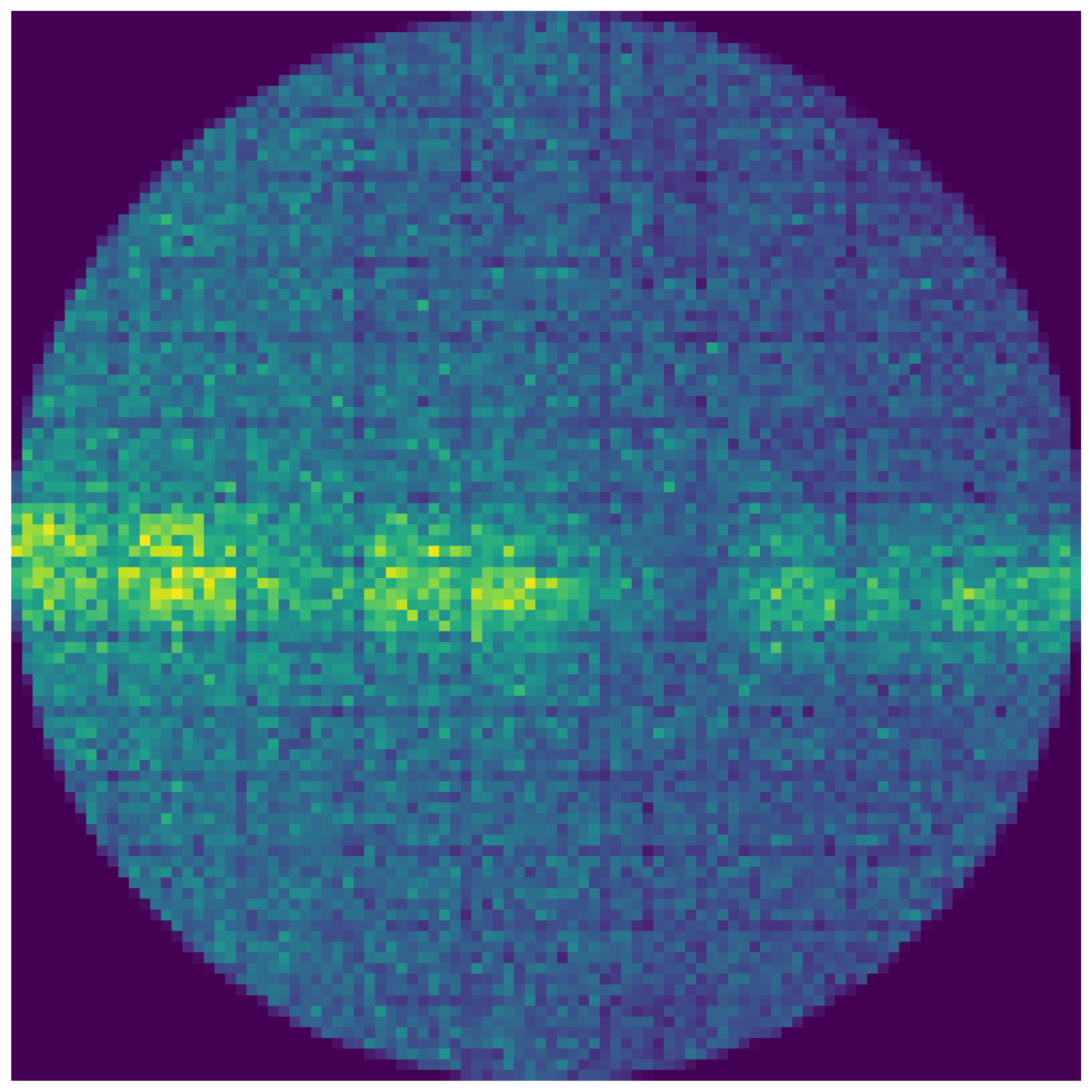}
      \put(3,150){\color{white}\textbf{(d)}}
    \end{overpic}
  \end{minipage}
  \hspace{0.01\textwidth}
  \begin{minipage}{0.32\textwidth}
    \centering
    \begin{overpic}[width=\linewidth]{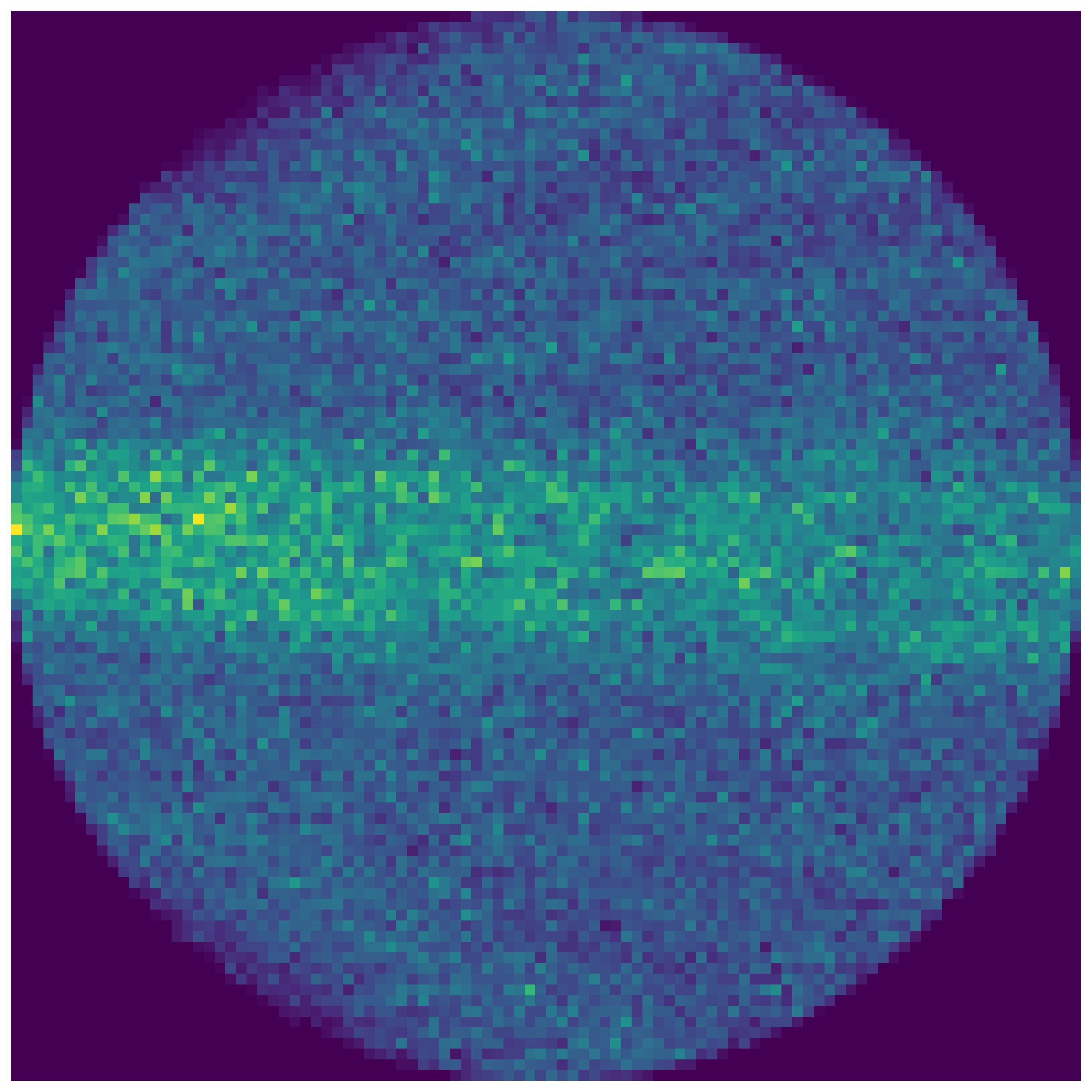}
      \put(3,150){\color{white}\textbf{(e)}}
    \end{overpic}
  \end{minipage}

  \caption{Comparison of experimental and simulated detector maps for the Ni[111] system, highlighting the effects of roll-up-induced directional perturbations on detector pattern formation. The detector maps for \textbf{a})~experimental data showing artefacts including enhanced and depleted zone lines and depleted poles, \textbf{b})~simulation data without the roll-up perturbation, and \textbf{c})~simulation data with the optimal roll-up perturbation energy distribution.
  We concentrate on a specific area (red circle) and zoomed-in views are shown for \textbf{d})~experimental data and \textbf{e})~simulation data with perturbations.
  }
  \label{fig:ni_figure_layout}
\end{figure*}

\begin{figure*}[htbp]
  \centering
 \begin{minipage}[t]{0.4\textwidth}
    \centering
    \begin{overpic}[height=6cm]{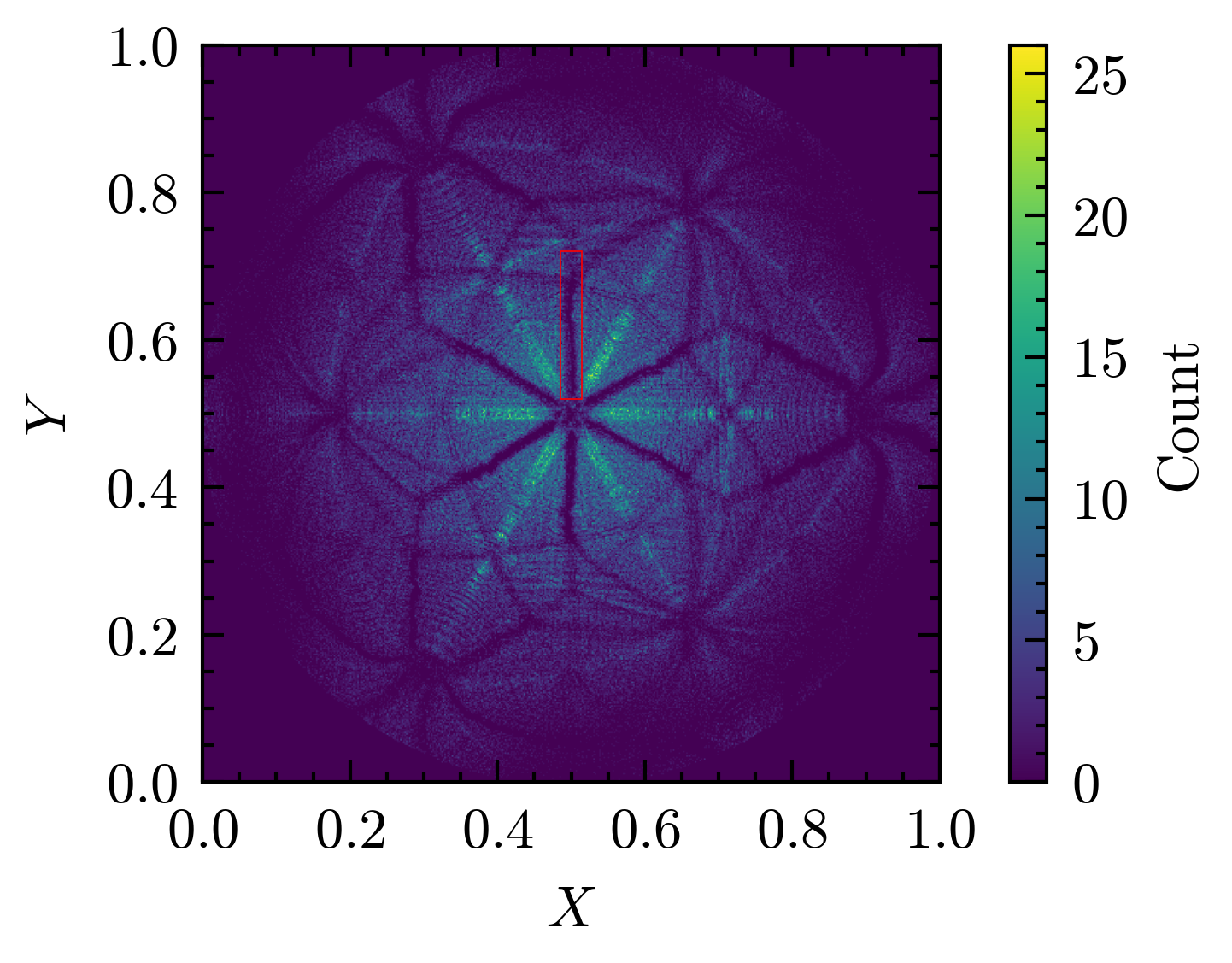}
      \put(2,170){\textbf{(a)}}
    \end{overpic}
  \end{minipage}%
  \hfill
  \begin{minipage}[t]{0.58\textwidth}
    \centering
    \begin{overpic}[height=6cm]{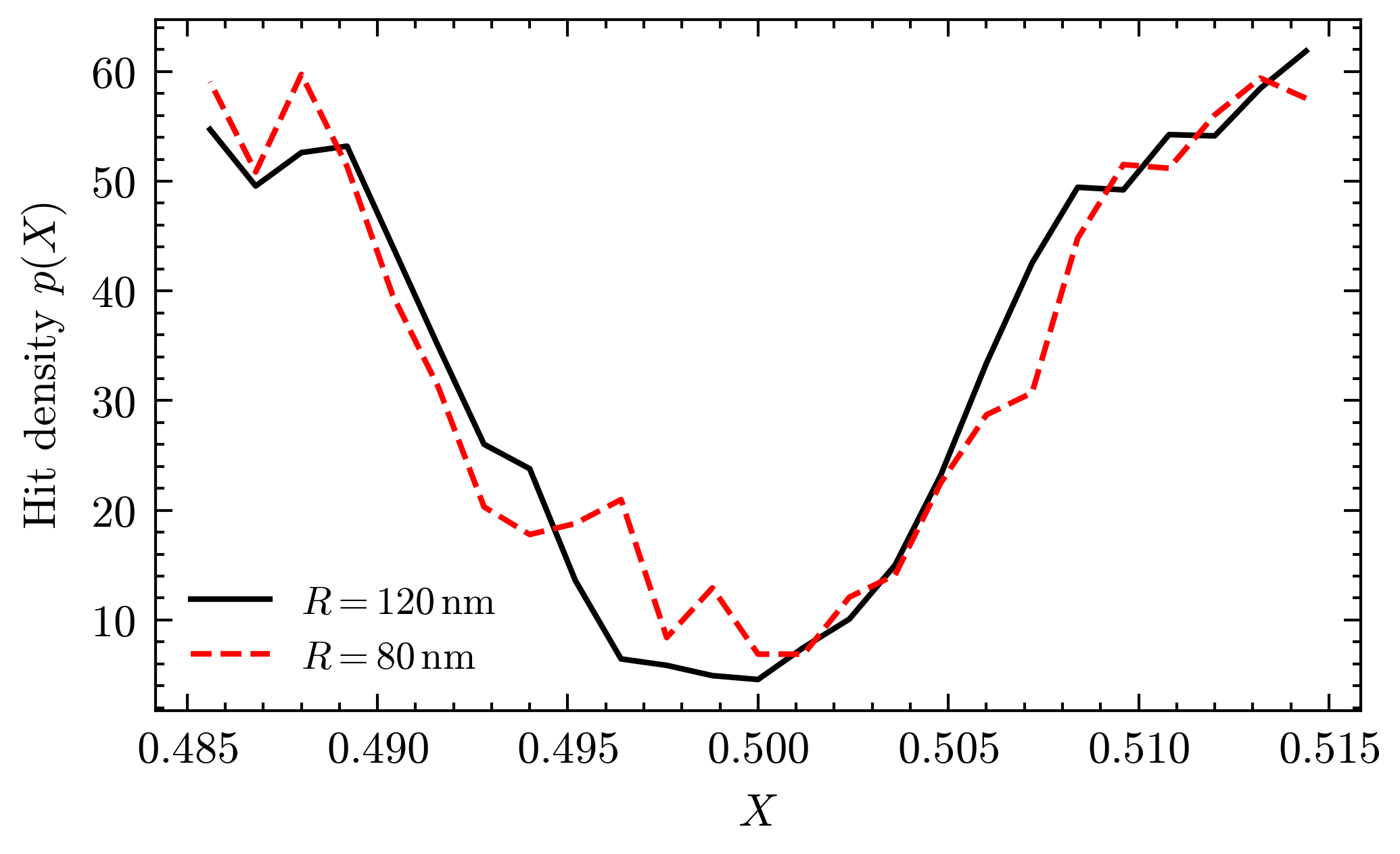}
      \put(2,170){\textbf{(b)}}
    \end{overpic}
  \end{minipage}
\vspace{1em}
 \begin{minipage}[t]{0.4\textwidth}
    \centering
    \begin{overpic}[height=6cm]{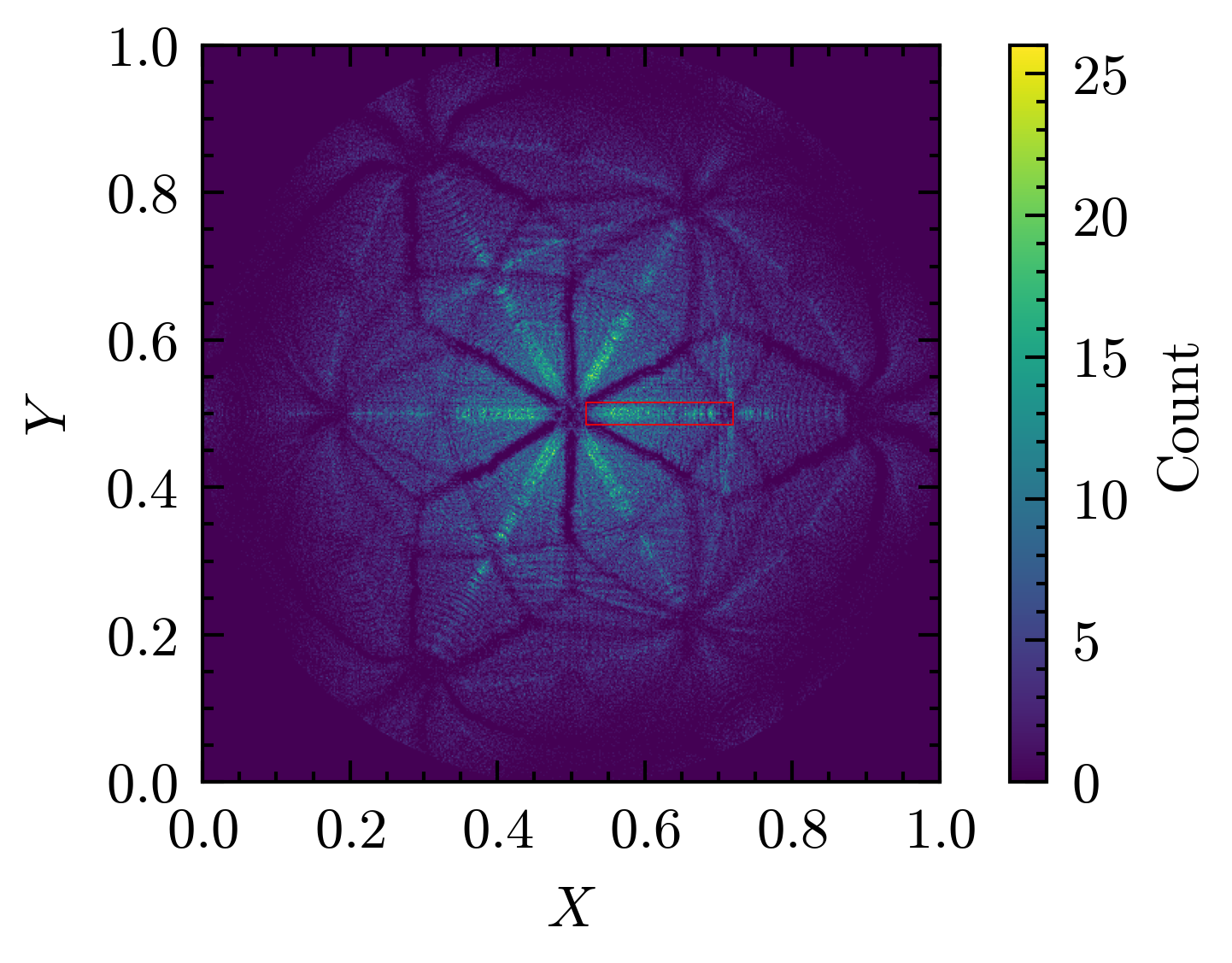}
      \put(2,170){\textbf{(c)}}
    \end{overpic}
  \end{minipage}%
  \hfill
  \begin{minipage}[t]{0.58\textwidth}
    \centering
    \begin{overpic}[height=6cm]{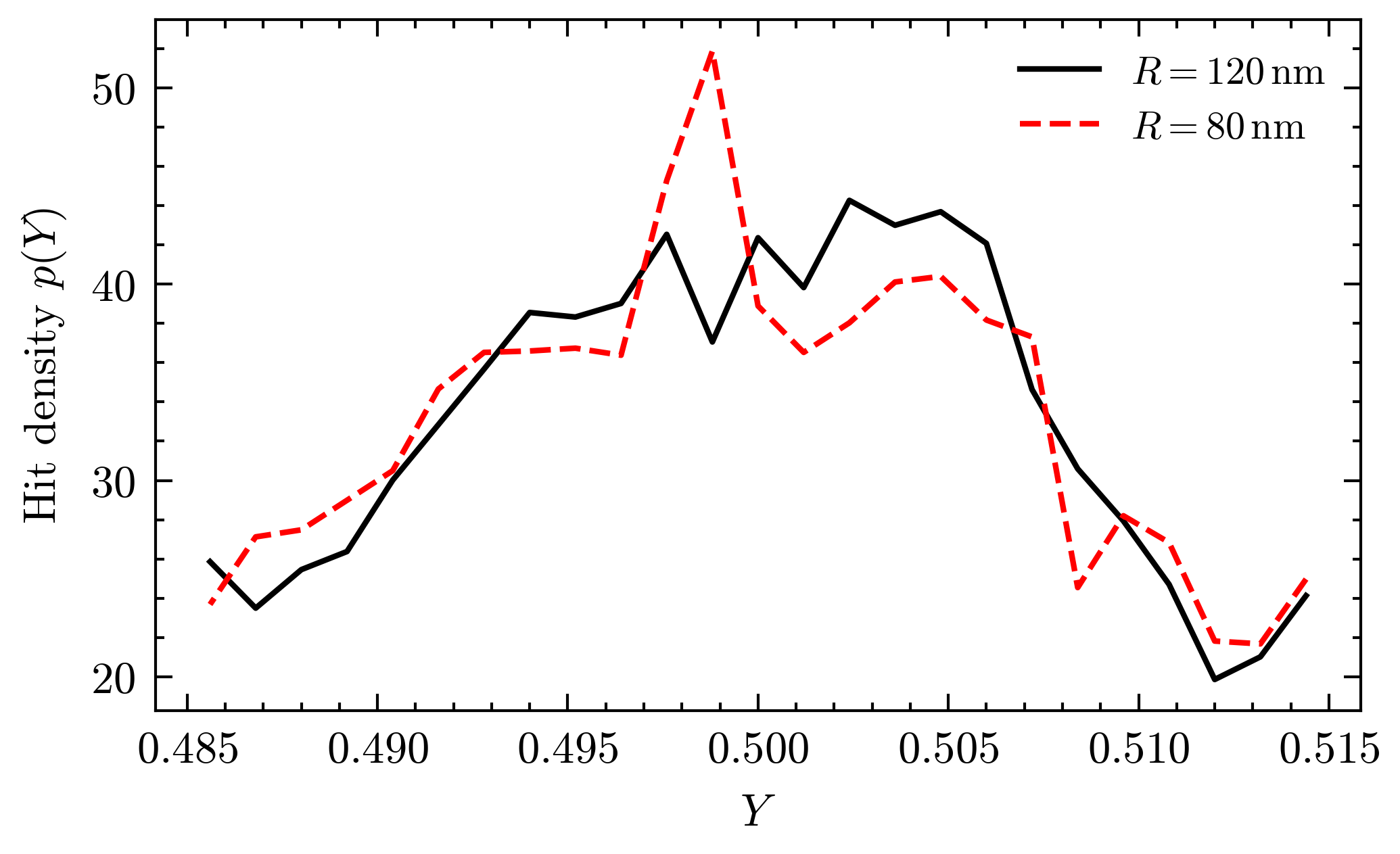}
      \put(2,170){\textbf{(d)}}
    \end{overpic}
  \end{minipage}
\vspace{1em}
  \caption{To illustrate the effect of specimen radius on the width of zone axes we take \textbf{(a)} a small rectangular section of the detector map around the depleted zone axis and compute \textbf{(b)} the normalized hit density $p(X)$ (averaged over the vertical $Y$ axis) for each horizontal bin for specimens with radii $R = 120$~nm and $R = 80$~nm.
  Similarly, we take \textbf{(c)} a small rectangular section of the detector map around the enhanced zone axis and compute \textbf{(d)} the normalized hit density $p(Y)$ (averaged over the horizontal $X$ axis) for each vertical bin for the two specimens.
  }
  \label{fig:system-comparison}
\end{figure*}

\subsection{Detector maps}

Although the SSIMs yield a quantifiable estimate of the similarity between the experimental and simulated detector maps, the actual detector maps have characteristic patterns that depend on factors such as the material crystallography and external experimental parameters (e.g., applied voltage, temperature). These patterns typically include enhanced and depleted zone lines, depleted poles, and concentric ring-like features surrounding the poles \cite{qi2022origin}, which are associated with atomic terraces on the specimen surface. 
These enhanced and depleted lines arise from trajectory focusing and defocusing near low-index crystallographic directions~\citep{vurpillot1999shape}, where the local electric field gradients are strongest. At such directions, ions emitted from neighboring terraces are preferentially steered toward or away from the zone axis, producing a enhanced or reduced hit density on the detector.
Understanding the origin of these detector map features is critical, as they represent artefacts that can influence the fidelity of the specimen reconstruction. 

The experimental detector map for Al[012] is shown in Fig.~\ref{fig:al_figure_layout}a. It shows a characteristic network of depleted lines (dark lines in the images), and depleted regions around poles. The simulated detector map for Al[012] in the absence of any perturbation ($E_\perp = 0$) cropped to the same field of view is shown in Fig.~\ref{fig:al_figure_layout}b. 
The depleted lines are much thicker, with narrow enhanced lines in the middle of them (which is absent in the experimental data). Similar features, such as rings around the poles, are seen, but they are different in appearance. These rings also encompass a much larger portion of the detector hits, making the areas far from the depleted lines also seem depleted of hits.
With the optimal perturbation energy distribution determined based on the SSIM (see Fig.~\ref{fig:al_figure_layout}c), the simulated detector map is much closer to the experimental one (Fig.~\ref{fig:al_figure_layout}a).
A magnified view of a selected region (shown by a red circle in Figs.~\ref{fig:al_figure_layout}a and~c) from the experimental and perturbed simulation maps is shown in Figs.~\ref{fig:al_figure_layout}d and e, respectively. Key features---such as the enhanced and depleted zone lines, and the depleted pole---are clearly reproduced in the perturbed simulation. The only clear difference is the slightly wider appearance of the depleted lines. 
The width of the widest depleted lines is only weakly affected by the magnitude of the applied perturbation. Their width is dominated not by the stochastic perturbations, but by the electrostatic roughness of the emitting surface. In the RRM, the local field landscape is determined by a classical surface-charge solution with an abrupt boundary (no field penetration) and a fixed hemispherical curvature. This construction tends to produce steep lateral field gradients near terraces, and the resulting defocusing leads to broad depleted lines whose amplitude scales essentially with the aspect ratio between the local roughness and the overall tip radius. Once these electrostatic gradients are set by the geometry and by the RRM approximation itself, adding small isotropic perturbations affects mainly the fine structure of the map but has only a minor influence on the intrinsic electrostatic width of the strongest depleted features.
This is also consistent with the fact that the simulated Al tip radius differs slightly from the experimental radius, which further affects the magnitude of these field gradients.

The same analysis can be performed for the roll-up case, compared with the experiments on Ni. The experimental detector maps for Ni[111] are seen in Fig.~\ref{fig:ni_figure_layout}a. They show very clear enhanced lines around a single pole and some depleted lines further away from the pole.
In comparison, the simulated detector map for Ni[111] in the absence of roll-up perturbation and associated lateral velocity component ($E_\perp = 0$), cropped to the same field of view, is seen in Fig.~\ref{fig:ni_figure_layout}b. 
In the unperturbed simulations the same enhanced lines are seen, but less prominently. The regions far from the lines then have a larger hit count than in the experiments. Between the enhanced lines, very strong depleted lines are seen.
With the optimal perturbation energy distribution determined based on the SSIM (Fig.~\ref{fig:ni_figure_layout}c), the detector map again looks much closer to the experimental one (Fig.~\ref{fig:ni_figure_layout}a). The enhanced lines are much more prominent and the hit count around them looks much closer to the experimental results. The clearest discrepancy is the appearance of even stronger depleted lines between the enhanced ones.
A magnified view of a selected region from the experimental and perturbed simulation maps (area indicated by the red circle in Figs.~\ref{fig:ni_figure_layout}a and c) is shown in Figs.~\ref{fig:ni_figure_layout}d and e, respectively. One can again see the key features---such as the enhanced zone lines and the depleted pole---reproduced in the perturbed simulation.\\

Some of the remaining discrepancy (more pronounced depleted and enhanced zone lines) can be attributed to our simplified roll-up implementation, which applies roll-up to all atoms. This likely overemphasizes local distortion effects. In future work, we aim to develop more selective roll-up criteria, potentially guided by molecular dynamics models (such as TAPSim-MD~\cite{qi2022ab}), to suppress unphysical depletions while preserving the enhanced features. The depleted lines themselves may serve as a useful indicator in tuning such criteria.
We also note that the absence of pronounced radial ring-like patterns in the simulated maps might be due to a cropping artefact: in order to compare directly with experimental detector maps, we cropped the simulation output to match the experimental field of view, which can reduce these radially symmetric features.
Finally, it is also important to note that in all of the analysis of the detector maps, the alignment of the experimental and simulation data is crucial. The detector maps have indeed been carefully aligned and cropped, but still some imperfections in the alignment can be seen. These are due to the experimental limitations (e.g. tilt of the sample) that are very hard to replicate in the simulations.

\subsection{System size effects}
As evident in Figs.~\ref{fig:al_figure_layout} and \ref{fig:ni_figure_layout}, simulated detector maps reveal zone lines (enhanced and depleted lines) that are noticeably wider than those observed experimentally. 
This discrepancy is, in part, a consequence of the simplifying electrostatic approximations inherent in the RRM, as discussed earlier. %Specifically, RRM assumes no field penetration into the specimen and imposes an abrupt discontinuous surface charge distribution, resulting in artificially steep electric field gradients at the specimen surface. These steep gradients amplify ion trajectory dispersion, particularly around low-index crystallographic directions, thereby producing wider zone-line features in the simulated detector maps.

However, another possible reason for the zone line enhancement would be the smaller size of the simulated sample, compared with the experimental sizes. The smaller sample naturally results in a higher radius of curvature for the specimen surface, and might therefore enhance the effects of the ion emission perturbations.
To assess the influence of specimen geometry on field evaporation behavior, we examined the effect of increasing the specimen tip radius on the width of depleted and enhanced zone axis lines in the detector maps.
Simulations were performed for the roll-up case pertaining to Ni, with the optimal perturbation energy distribution, and with specimens of radius 120~$a$, in comparison with the previous radius of 80~$a$.
One can then compare the detector maps around the depleted zone lines (area enclosed in red in Fig.~\ref{fig:system-comparison}a), averaging over the vertical axis, and see how this average hit density $p(X)$ over the zone line changes with the system size (Fig.~\ref{fig:system-comparison}b). One can clearly see that there is almost no difference.
Same procedure can be followed over the enchanced line (see Fig.~\ref{fig:system-comparison}c), and the resulting hit density (averaged over the horizontal coordinate~$X$) $p(Y)$ (Fig.~\ref{fig:system-comparison}d) again shows practically no difference between the two system sizes.
The results clearly show that increasing the specimen radius has negligible effect on the width of zone axis features, both for depleted and enhanced lines. This suggests that other factors---such as the electrostatic approximation---are more critical in determining these detector patterns.

\section{Conclusions}
In this work we demonstrated that incorporating minimal, physically motivated and system-specific, stochastic emission mechanisms---namely lateral velocity perturbations and roll-up motion---into field evaporation simulations substantially improves the agreement between simulated and experimental detector maps for both Al and Ni, i.e.~weakly and strongly bonded systems. 
These mechanisms reproduce the characteristic artefacts observed in experiments and thus capture the physical processes better than previous APT models.
The remaining discrepancies are attributed to simplifications in the electrostatic formulation of the simulation model, as well as experimental imperfections related to e.g.~the alignment of the specimen. 

To our knowledge, out of the state-of-the-art APT simulations only MD-based methods such as TAPSim-MD~\citep{qi2022ab} have achieved comparable realism. These methods are currently limited to much smaller specimen sizes and are computationally infeasible for large-scale use. In contrast, our model is here shown to be applicable to the tip sizes comparable to the experimental ones, and at this level, is the first approach to replicate this level of experimental agreement.\\

In our analysis, the best match with experiments is obtained with stochastic perturbation energies with fairly high mean values and high dispersion.
While this is consistent also with some previous experimental observations, the energies exceed the simple lower-bound estimates from field-induced polarization.
More detailed molecular dynamics or similar simulations are needed to explain the fundamental physics behind these stochastic perturbation energy distributions, and possibly make the shape of the energy distributions even more accurate.
Such simulations could also assist in informing criteria for roll-up based on local geometry, field, or bonding, to be used in the RRM instead of roll-up of all atoms used in this work.

Additionally, a full microscopic description of field evaporation would require resolving complex electronic and electrostatic processes during ion detachment, which is beyond the scope of the present work. Instead, we adopt a coarse-grained description that focuses on the effective lateral impulse imparted to the ion at emission, as this is the quantity that directly determines detector hit positions.
We note that the Robin--Rolland model itself relies on similar approximations. In particular, as already discussed by Ref.~\cite{rolland2015meshless}, classical image-charge interactions and dynamic surface charge redistribution are neglected, as their explicit treatment would be computationally prohibitive. While such effects may influence the detailed trajectory dynamics, they are not expected to generate dominant lateral forces. Our aim here is therefore limited to identifying and modeling the effective lateral perturbations required to reproduce the main experimental features of detector maps. More complete microscopic treatments remain an important direction for future work.\\

By significantly narrowing the gap between experimental observations and large-scale field evaporation simulations, our approach provides a robust foundation for the development and validation of next-generation reconstruction algorithms. Current reconstruction algorithms rely on several geometric assumptions which are violated when lateral perturbations and roll-up events occur. Our results identify the stochastic perturbations needed to best reproduce experimental detector maps. These quantitative perturbation models can then be directly incorporated into modified reconstruction algorithms (e.g.,~dynamic magnification models~\cite{gault2011dynamic, hatzoglou2019enhanced, hatzoglou2025effect}, Bayesian inference, trajectory-corrected back-projection~\cite{haley2011atom}).
This is essential for enhancing the spatial fidelity of reconstructions and for improving the reliability of quantitative compositional analysis in APT. 
Extending the framework to include additional physical effects---such as surface topography evolution, correlated evaporation events, and laser-induced thermal excitations---will further strengthen the predictive power of APT.

\begin{acknowledgements}

I.L. and A.S acknowledge the funding from Academy of Finland (341440 and 346603).
M.J.A. acknowledges support from the Academy of Finland (Center of Excellence program, 278367 and 317464).
T.M. and M.A. acknowledge support from the FinnCERES flagship (151830423), Business Finland (211835, 211909, 211989, and 210129), the Research Council of Finland (13361245), and Future Makers programs.
The authors acknowledge the computational resources provided by the Aalto University School of Science “Science-IT” project. 

\end{acknowledgements}

\section*{Data availability}
The data that support the findings of this article are openly available~\cite{zenodo}.
% The data that support the findings of this article are not publicly available. The data are available from the authors upon reasonable request.

\appendix

\section{Electrostatic formulation of the Robin--Rolland model}
\label{app:robin}

In classical electrostatics, an ideal conductor held at a fixed potential forms an equipotential surface enclosing a region where the electric field is zero. Free electrons redistribute until this equilibrium is reached, causing the volume charge density inside the conductor to vanish while a non-zero surface charge density develops on its boundary. 

For a smooth conducting surface $S$, the surface charge density $\sigma$ at a point $P \in S$ can be expressed using Robin's integral equation~\cite{robin1886distribution}
\begin{equation}
\label{eq:Robin}
\sigma(P) = \frac{1}{2\pi} 
\iint_S 
\frac{\hat{\bm{n}}_P \times \bm{u}}{|\bm{u}|^3} \, 
\sigma(P') \, dS',
\end{equation}
where $P$ and $P'$ are points on the surface $S$, $\hat{\bm{n}}_P$ is the outward unit normal to the surface at point $P$, $\bm{u} = \overrightarrow{P'P}$ is the vector from $P'$ to $P$, and $dS'$ is an infinitesimal surface element at $P'$. 

Robin also proposed an iterative procedure for evaluating Eq.~\ref{eq:Robin}. Let $\{f_n(P)\}$ be a sequence of functions defined on $S$ such that
\begin{equation}
\label{eq:Robin-iter}
f_{n+1}(P) = \frac{1}{2\pi} 
\iint_{S} 
\frac{\hat{\bm{n}}_P \times \bm{u}}{|\bm{u}|^3} \,
f_n(P') \, dS'.
\end{equation}
Here, $f_0$ denotes an arbitrary initial function on $S$. Under the assumption that the surface $S$ is convex, Robin demonstrated that this iterative sequence converges to the surface charge density $\sigma(P)$.

In classical treatments, this surface charge is considered to lie on a smooth continuous surface of infinitesimal thickness. In real materials, however, charge tends to accumulate at the positions of atoms on the surface. These atoms may be treated as partial ions, with protruding atoms carrying more charge and therefore producing a stronger local electric field. In the RRM~\cite{rolland2015meshless} used in this work, the continuous surface $S$ is discretized onto the positions of the emitter surface atoms, enabling the equilibrium surface charge distribution to be computed directly from the surface-atom geometry.

Each surface atom is assigned a small surface area $s_{\text{at}}$. When a constant electrostatic potential is applied, each surface atom acquires a surface charge density $\sigma_i$, and the corresponding charge on atom $i$ is $q_i = s_{\text{at}}\,\sigma_i$. The surface charge can then be computed using a discretized form of Robin's equation,
\begin{equation}
\label{eq:surfaceCharge_app}
\frac{q_i}{s_{\text{at}}} 
= \frac{1}{2\pi} 
\sum_{\substack{k=1 \\ k \ne i}}^{N} 
q_k 
\frac{\hat{\bm{n}}_i \times \bm{r}_{ik}}{|\bm{r}_{ik}|^3},
\end{equation}
where $\hat{\bm{n}}_i$ is the local outward normal, $\bm{r}_{ik} = \bm{r}_i - \bm{r}_k$, and the index $k$ runs over all $N$ surface atoms. The iterative scheme used in the main text,
\begin{equation}
\label{eq:disc_iter_app}
\frac{q_{i,n+1}}{s_{\text{at}}} 
= \frac{1}{2\pi} 
\sum_{\substack{k=1 \\ k \ne i}}^{N} 
q_{k,n} 
\frac{\hat{\bm{n}}_i \times \bm{r}_{ik}}{|\bm{r}_{ik}|^3},
\end{equation}
is directly analogous to Eq.~\ref{eq:Robin-iter} and converges to the discrete equilibrium charge distribution for convex surfaces~\cite{rolland2015meshless}.

A useful property of the continuous formulation is charge conservation during the iterative process. By noting that the surface viewed from point $P$ subtends a solid angle of $2\pi$, one finds
\begin{equation}
\label{eq:solid_angle}
\iint_S f_{n+1}(P) \, dS 
= \iint_{S} f_n(P') \, dS'.
\end{equation}
It follows that the total charge carried by the conductor remains constant throughout the iterations, i.e.,
\begin{equation}
\iint_S \sigma_{n+1}(P) \, dS = \iint_S \sigma_n(P) \, dS.
\end{equation}
To ensure the same behavior for the discrete scheme in Eq.~\ref{eq:disc_iter_app}, the atomic surface area $s_{\text{at}}$ is recomputed at each iteration so that
\begin{equation}
\label{eq:charge_conservation_app}
\sum_{i=1}^{N} q_{i,n+1} 
= \sum_{i=1}^{N} q_{i,n} .
\end{equation}
This rescaling preserves the total charge of the emitter while allowing the individual $q_i$ to relax toward their equilibrium values.

\section{Field-induced lateral energy via polarization forces} \label{app:polar}
Here we show a simple calculation for the energy scales associated with field-induced polarization.
The purpose of this appendix is not to compute this force for a specific geometry---a task that requires an atomistically resolved electrostatic model---but to provide a simple estimate for the energy associated with tangential polarization forces. This estimate is used only as a rough guide for the perturbation energies involved.
A neutral atom near a sharp field emitter experiences a spatially varying electric field $\bm{E}$, which induces a dipole moment $\bm{p} = \alpha \bm{E}$, where $\alpha$ is the atomic polarizability. The corresponding polarization energy is
\begin{equation}
U = -\frac{1}{2} \alpha \|\bm{E}\|^2,
\end{equation}
with the prefactor 1/2 accounting for the induced nature of the dipole. The force resulting from field inhomogeneity is
\begin{equation}
\bm{F} = -\nabla U = \alpha \|\bm{E}\| \nabla \|\bm{E}\|,
\end{equation}
which drives atoms toward regions of higher field strength.

%For small displacements $\Delta r$ along the tip surface, the work done by this force---i.e., %energy due to polarization---is approximately
%\begin{equation}
%E_{\mathrm{pol}} \approx \alpha E \frac{\mathrm{d} E}{\mathrm{d} r} \Delta r.
%\label{eq:rollup_general}
%\end{equation}
Only the tangential component of this force contributes to lateral motion along the surface. For a small lateral displacement $\Delta r$, the work done by this force---i.e., energy due to polarization---is approximately
\begin{equation}
E_{\mathrm{pol}} \approx \alpha \|\bm{E}\| \frac{\mathrm{d}\|\bm{E}\|}{\mathrm{d}r} \Delta r
\label{eq:rollup_general}
\end{equation}
where $\frac{\mathrm{d}\|\bm{E}\|}{\mathrm{d}r}$ denotes the field gradient along the surface.
This is a general geometry-independent expression and obtaining a precise numerical value for $\frac{\mathrm{d}\|\bm{E}\|}{\mathrm{d}r}$ would require knowledge of the detailed atomic-scale curvature landscape (terraces, steps, microfacets) and of the height of the atom above the surface, because tangential field gradients decay rapidly with distance from the emitter. For this reason, any simple analytical substitution for~$\frac{\mathrm{d}\|\bm{E}\|}{\mathrm{d}r}$ produces only a lower bound on the energy scale.

To obtain an order-of-magnitude estimate, we consider a simplifying assumption $\frac{\mathrm{d}\|\bm{E}\|}{\mathrm{d}r} \approx \frac{\mathrm{d}\|\bm{E}\|}{\mathrm{d}R}$, i.e. that the polarization is just due to a roll-up-like displacement in the normal direction.
We then estimate this gradient by assuming a conical tip at voltage $V$, the field near the apex behaving as $\|\bm{E}\| \approx V / (k R)$, where $R$ is the radius of curvature and $k$ depends on geometry (typically of the order 1-10). This yields
\begin{equation}
E_{\mathrm{pol}} \approx \alpha \frac{V^2}{k^2} \frac{\Delta r}{r^3}.
\label{eq:rollup_energy}
\end{equation}

%Assuming a conical tip at voltage $V$, the field near the apex scales as $E(r) \approx V / (k r)$, where $r$ is the local radius of curvature and $k$ depends on geometry (typically of the order 1-2). Differentiation yields $\mathrm{d}E/\mathrm{d}r = -V / (k r^2)$, and Eq.~\eqref{eq:rollup_general} becomes
%\begin{equation}
%E_{\mathrm{pol}} \approx \alpha \frac{V^2}{k^2} \frac{\Delta r}{r^3}.
%\label{eq:rollup_energy}
%\end{equation}

This expression quantifies the energy gained by an atom during lateral motion under the influence of polarization forces. Upon field evaporation, this energy is converted to lateral kinetic energy,
\begin{equation}
E_{\mathrm{pol}} = E_\perp = \frac{1}{2} m v_\perp^2,
\end{equation}
which introduces a transverse velocity component in the ion trajectory.
For parameters matching the simulations for Ni ($V = 1000$~V, $k = 1$, tip radius $r = 80a = 28$~nm, %\TM{Our tip radius in the simulations is $80a$ and $a = 3.52$~\AA, so $r = 28$~nm.},
step length $\Delta r = a = 0.35$~nm, $\alpha_{\text{Ni}} = 6.8 \times 10^{-40}$~Fm$^2$), Eq.~\eqref{eq:rollup_energy} yields
\begin{equation}
E_{\mathrm{pol}} \approx 0.068 \, \mathrm{eV},
\end{equation}
and for Al (with $\Delta r = a = 0.40$~nm and $\alpha_{\text{Al}} = 9.8 \times 10^{-40}$~Fm$^2$) %\TM{Same here, $a = 4.04$~\AA, so $r = 32$~nm.}
\begin{equation}
E_{\mathrm{pol}} \approx 0.075 \, \mathrm{eV}.
\end{equation}
The precise values depend on the local surface configuration and these estimates should be interpreted as lower bounds due to the rapidly decaying nature of lateral field gradients and the simplified treatment of the near-surface geometry.

%These estimates are significantly smaller than the mean perturbation energies ($\sim 0.15–0.3~\mathrm{eV}$) extracted by matching simulated and experimental detector maps. This comparison supports the conclusion that polarization alone cannot account for the full magnitude of the observed lateral perturbations and that additional mechanisms---in particular roll-up-related bond-breaking asymmetry and local curvature effects---must contribute to the effective lateral momentum imparted at emission.

\bibliography{references} % Produces the bibliography via BibTeX.

\end{document}